\documentclass[letterpaper,titlepage,11pt]{article}
\usepackage{amsmath,amssymb,mathtools,bm}
\usepackage{physics}
\usepackage{cancel}
\usepackage{slashed}
\usepackage{mathrsfs}
\usepackage{color}
\usepackage[svgnames]{xcolor}
\usepackage{enumerate}
\usepackage{graphicx,booktabs,array}
\usepackage{tikz}
\usepackage{caption}
\usepackage[symbol]{footmisc}
\usepackage{enumitem}

\usepackage{multirow}
\usepackage{booktabs}

\usetikzlibrary{arrows.meta, positioning, calc, decorations.pathmorphing, shapes.geometric}
\usepackage{tikzsymbols}
\usepackage{fontawesome5}
\usepackage{pifont}
\usetikzlibrary{shadows}

\usepackage[numbers,sort&compress]{natbib}
\allowdisplaybreaks

\usepackage[
colorlinks=true,
linkcolor=DarkBlue,
anchorcolor=Black,
urlcolor=DarkRed,
filecolor=green,
citecolor=DarkRed,
linktoc=page,
pdfstartview=FitV,
pdftitle={Classical and Stochastic Trans-Planckian Censorship},
pdfauthor={Omer Guleryuz},
pdfsubject={},
pdfkeywords={},
bookmarksopen=true
]{hyperref}

\usepackage[nameinlink,capitalise]{cleveref}

\renewcommand{\dd}{\mathrm d}

\newcommand{\Mp}{M_{\rm Pl}}
\newcommand{\PZ}{\mathcal{P}_\zeta}
\newcommand{\cN}{\mathcal{N}}

\begin{document}

 \hypersetup{pageanchor=false}
 \begin{titlepage}
			
			\bigskip
			
			\begin{center}
				{\LARGE\bfseries Classical and Stochastic Trans-Planckian Censorship}
				\\[10mm]
				\textbf{Omer Guleryuz}\\[5mm]
				\vskip 25pt

				{\em \hskip -.1truecm Department of Physics, Istanbul Technical University, \\
					Maslak 34469 Istanbul, Türkiye \vskip 5pt }

				{{\tt \href{mailto:omerguleryuz@itu.edu.tr}{omerguleryuz@itu.edu.tr}}
 
 }

			\end{center}
			
			\vspace{3ex}

			\begin{center}
				{\bfseries Abstract}
			\end{center}
\begin{quotation}
The trans-Planckian censorship conjecture (TCC) limits inflation to a deadline of $N_{\rm TCC}=\ln(\Mp/H_{\rm end})$ e-folds. We formulate it in stochastic inflation as a bound at confidence $1-\delta$ on the probability that a history outlasts the deadline, with three readings: one worldline, the region grown from one Hubble patch, and the reheated volume. Each reading becomes a closed-form condition, rigorous for a linear roll in leading-order stochastic inflation (for the region, in a branching model) and sharp at leading order in $N_{\rm TCC}$: a linear roll whose classical history lasts $N_{\rm cl}$ e-folds respects the TCC if $1-N_{\rm cl}/N_{\rm TCC}\geq\{2\PZ[w+\ln(1/\delta)/N_{\rm TCC}]\}^{1/2}$, with $w=0$ for a worldline and $w=3$ for a region. On a linear slope without vacuum decay, the worldline and region readings are never weaker than the classical TCC, and on the CMB-normalized attractor all readings agree closely with it, so the bounds on the scale of inflation stand. A classical phase that ends in a flat quantum well, which the overdamped classical TCC excludes, is allowed with a margin of a few e-folds per worldline. Near the onset of eternal inflation, the readings differ. On a linear slope, a region must leave the fraction $\sqrt{6\PZ}$ of the deadline unused at leading order: eternal inflation is excluded for every classical history that uses a finite fraction of the deadline, as is a band of non-eternal slopes, and the slope bound becomes $|V'|>V^{3/2}/[\sqrt2\,\pi(1-N_{\rm cl}/N_{\rm TCC})]$. Per worldline, every slope is allowed, given a margin of a few standard deviations. A false vacuum must decay at about the eternal threshold $\Gamma=9H^4/(4\pi)$ or faster for a region or the volume, but only faster than $3H^4\ln(1/\delta)/(4\pi N_{\rm TCC})$ per worldline. Whether the TCC forbids eternal inflation depends on the reading.
\end{quotation}
			
			\vfill
			
 \begin{center}{\flushleft{\today}}\end{center}
		\end{titlepage}
		\hypersetup{pageanchor=true}
		\setcounter{page}{1}
		\tableofcontents

\newpage

\section{Introduction}\label{sec:intro}
The trans-Planckian censorship conjecture (TCC) states that no fluctuation whose wavelength was ever shorter than the Planck length may cross the Hubble horizon and freeze into a classical perturbation~\cite{Bedroya:2019snp}. For an expanding phase, it limits the number of e-folds to a deadline,
\begin{equation}\label{eq:intro-tcc}
N<N_{\rm TCC}\equiv\ln\frac{\Mp}{H_{\rm end}}=\frac12\ln\frac{S_{\rm end}}{8\pi^2}\,,
\end{equation}
where $S_{\rm end}$ is the Gibbons--Hawking entropy at the end of inflation~\cite{Gibbons:1977mu}. The best-known consequences of the TCC are an upper bound on the scale of inflation, $V^{1/4}<6\times10^8\,$GeV with a tensor-to-scalar ratio $r<10^{-30}$~\cite{Bedroya:2019tba}, a bound whose precise value depends on the assumed history around inflation~\cite{Mizuno:2019bxy,Kamali:2019gzr,Brandenberger:2019eni}, a lifetime bound for metastable de Sitter vacua~\cite{Bedroya:2019snp}, and the statement that it ``marginally'' rules out eternal inflation in rolling potentials dual to cascades of de Sitter bubbles~\cite{Bedroya:2020rac}. Eternal inflation~\cite{Vilenkin:1983xq,Linde:1986fd} had earlier been confronted with the de Sitter swampland conjectures~\cite{Obied:2018sgi,Ooguri:2018wrx} in Refs.~\cite{Matsui:2018bsy,Dimopoulos:2018upl,Kinney:2018kew,Brahma:2019iyy,Wang:2019eym,Rudelius:2019cfh,Blanco-Pillado:2019tdf}. The TCC sharpens the trans-Planckian problem of inflation~\cite{Martin:2000xs} and has been related to de Sitter complementarity, horizon chaos and holography~\cite{Bedroya:2020rmd,Aalsma:2020aib,Bedroya:2022tbh,Bedroya:2025fie,Bedroya:2024zta}; its status as a swampland condition is debated~\cite{Saito:2019tkc,Burgess:2020nec,Dvali:2020cgt}, and we treat it as a stated condition.

All of these statements concern a single classical history. In stochastic inflation~\cite{Starobinsky:1986fx,Starobinsky:1994bd} every Hubble patch follows its own history, and the duration of inflation is a first-passage time with a known distribution~\cite{Vennin:2015hra,Pattison:2017mbe,Ezquiaga:2019ftu}. The TCC then becomes a bound on the probability that a history outlasts the deadline, and it can be imposed on one worldline, on the whole region that grows from one Hubble patch, or on the reheated volume, a choice the classical theory never has to make. Two earlier works are closest to ours. Reference~\cite{Anchordoqui:2026edd} bounds the lifetime of quasi-de Sitter phases at any point of moduli space by the minimum of a classical-roll term and a stochastic-diffusion term, from the distance conjecture and a generalized Higuchi bound, at a fixed confidence level, and finds this bound weaker than the TCC. Reference~\cite{Guleryuz:2026rdz} studies the survival of stochastic histories near swampland boundaries and introduces a margin that keeps a chosen fraction of them, the analogue of the confidence level used here. In both, the bound rests on a distance in field space. Here the TCC itself is the bound, a deadline in time; drift and diffusion act together in one first-passage problem, and the three readings are compared.

We ask whether the stochastic TCC is stronger or weaker than the classical one, and whether it changes what the classical TCC allows or forbids. The answers follow from one inequality. A slow roll whose classical history lasts $N_{\rm cl}$ e-folds, with power spectrum $\PZ$, respects the TCC at confidence $1-\delta$ if the fraction of the deadline it leaves unused obeys
\begin{equation}\label{eq:intro-main}
1-\frac{N_{\rm cl}}{N_{\rm TCC}}\;\geq\;\sqrt{2\PZ\Big(w+\frac{\ln(1/\delta)}{N_{\rm TCC}}\Big)}\,,\qquad
w=\begin{cases}0&\text{worldline}\,,\\3&\text{region}\,,\end{cases}
\end{equation}
rigorously for a linear roll in leading-order stochastic inflation (Section~\ref{sec:stoch}). The classical TCC asks only that the left-hand side be positive. Per worldline, the stochastic TCC asks for a margin of a few standard deviations of the duration; for a region it asks for at least the fraction $\sqrt{6\PZ}$ of the deadline, as if the worldline condition were imposed on every Planck-sized cell of the initial patch. The inequality is sharp at leading order in $N_{\rm TCC}$ and settles both questions:
\begin{itemize}\setlength{\itemsep}{1pt}
\item On a linear slope without vacuum decay the worldline and region readings are, short of the flat limit, never weaker than the classical TCC, and neither is the volume reading, away from eternal inflation. On the attractor normalized to the CMB all readings agree with it to a fraction of an e-fold (Section~\ref{sec:slope-cmb}).
\item Near the onset of eternal inflation, $\PZ=1/6$~\cite{Creminelli:2008es}, the readings differ. The region reading
excludes eternal inflation for every classical history that uses a finite fraction of the deadline, fixes the
coefficient of the slope bound, and also excludes a band of slopes that are not eternal; the worldline reading allows every slope, given a margin of a few standard deviations of the duration (Section~\ref{sec:slope-eternal}). False vacua split the same way (Section~\ref{sec:metastable}).
\item A classical phase that ends in a flat quantum well, as in models of primordial black holes, is excluded outright by the classical TCC, since its overdamped classical duration diverges; the stochastic TCC asks instead for a margin of a few e-folds per worldline and, for a well below the eternal size, caps the classical phase at a fraction of the deadline for a region (Section~\ref{sec:wells}).
\end{itemize}
The TCC is motivated both by single-observer arguments and by a statement about all modes (Section~\ref{sec:stoch-which}), so we do not choose between the readings; whether it forbids eternal inflation depends on the reading. Section~\ref{sec:examples} applies the conditions to models, among them one that fits the CMB data, and shows where the classical TCC is too weak, where it is too strong, and where it gives no definite verdict.

The ingredients are standard: the first-passage statistics of the duration~\cite{Vennin:2015hra,Pattison:2017mbe,Ezquiaga:2019ftu}, the onset of eternal inflation and the tilt by the volume weight~\cite{Creminelli:2008es,Winitzki:2008ph,Animali:2024jiz}, tree models of eternal inflation~\cite{Harlow:2011az}, and the survival theory of branching Brownian motion with absorption~\cite{KESTEN19789,10.1214/ECP.v12-1259,berestycki2014critical,AHL_2022__5__921_0}. Two limits of the results are known: for a history that uses a vanishing fraction of the deadline, the slope coefficient is the onset value~\cite{Rudelius:2019cfh},
and at leading order the region and volume bounds on the decay rate of a false vacuum reduce to the eternal threshold~\cite{Guth:1982pn}. The new results are (i)~the formulation of the TCC as a deadline for a first passage with three readings, in which the region reading is the survival, up to the time $\ln(\Mp/H_{\rm end})$, of branching Brownian motion with absorption; (ii)~the inequality \eqref{eq:intro-main} and its volume counterpart, with the proof that they are sharp at leading order in $N_{\rm TCC}$ and the resulting order of the readings; and (iii)~their consequences: the robustness of the bounds on the scale of inflation, the excluded band of non-eternal slopes and the dependence of the slope coefficient on the fraction of the deadline used, the corrections to the bounds on the decay rate and the much weaker worldline bound, the margins for quantum wells, and the behavior of volume regulators, which range from violation for every model to a positive exponent for every slope. The inequality itself is elementary, a Chernoff bound combined with a first-moment bound; its value is that it is sharp, so that each reading of the TCC becomes a closed-form condition. The conformal diagrams of Fig.~\ref{fig:penrose}, drawn from a lattice simulation, illustrate the regimes and enter no derivation.

The inequalities for a linear roll, the order of the readings at leading order, the sharp threshold \eqref{eq:stoch-step} of the region reading, the criterion for eternal inflation and the formulas for a false vacuum whose patches decay independently are theorems or exact formulas, proved in the appendices or quoted from the probability literature; for the region they hold in the branching model of Section~\ref{sec:stoch-readings}. Derived at the level of rigor usual in physics are the Bernstein form for a varying power spectrum, the normal approximation on the attractor, the estimates of the eternal probability near the end surface,
the finite-history exponents and the necessity of the region condition for wells. The exact region readings at finite deadlines are numerical, and Section~\ref{sec:beyond} is heuristic.  We work with overdamped, leading-order stochastic inflation of a minimally coupled inflaton, with a constant decay rate where vacuum decay is included, and use it near the onset of eternal inflation as well, where the curvature perturbation is of order one, as the standard model of eternal inflation. Ultra-slow roll, quantum corrections to the stochastic description~\cite{Gorbenko:2019rza,Mirbabayi:2019qtx,Baumgart:2019clc,Cohen:2020php,Green:2026nnw} and the backreaction of the fluctuating geometry are left out. We use reduced Planck units, $\Mp=1$ unless written explicitly, and e-folds as time.

\section{Classical Censorship}\label{sec:classical}
\subsection{The deadline as a surface}\label{sec:classical-deadline}
\begin{figure}[htb]\centering
\includegraphics[width=\textwidth]{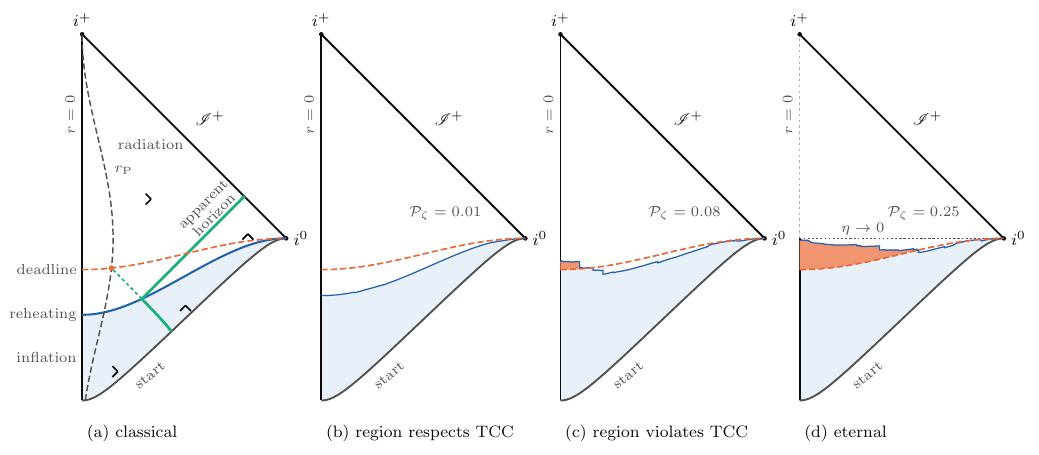}
\caption{Conformal diagrams of classical and stochastic censorship. The background is a flat FRW universe with quasi-de Sitter inflation followed by radiation (Appendix~\ref{app:lattice}). All panels use the toy values $N_{\rm cl}=1.5$ and, for the deadline (orange, dashed), $N_{\rm TCC}=2.5$, that is $H_{\rm end}=e^{-2.5}\Mp$, chosen for visibility; at such values the stochastic description is not quantitative. (a) Classical: the apparent horizon (green) with light-sheet wedges, the comoving sphere $r_{\rm P}$ (gray, dashed) that was Planck-sized at the start, and the deadline, where the inflationary apparent horizon, continued as a dotted green line, would reach $r_{\rm P}$ (orange dot). (b)--(d) Reheating times from one realization of stochastic inflation on a three-dimensional comoving lattice, with the same noise for three slopes (Appendix~\ref{app:lattice}), overlaid on the background diagram; the evolution after reheating is not simulated. Each diagram is centered on the point of the initial Hubble patch that reheats last; the blue curve is the reheating surface along one comoving ray from it, and orange marks the patches that inflate past the deadline. (b) $\PZ=0.01$: the whole region reheats before the deadline. (c) $\PZ=0.08$: the slope is not eternal and a typical point reheats in time, but the last patch of the region violates the TCC. (d) $\PZ=0.25$: eternal inflation; the subtree of the central cell never stops inflating, so the region reaches the future boundary $\eta\to0^-$ (dotted) above the center, where the axis $r=0$ is drawn faint; the single worldline followed along the ray still reheats, if late (Appendix~\ref{app:lattice}). For these parameters the region reading is violated with probability $10^{-7}$, $0.56$ and $0.82$ in (b)--(d), from Eq.~\eqref{eq:app-fkpp}.}
\label{fig:penrose}
\end{figure}

In a flat FRW universe with conformal time $\eta$ the apparent horizon is the comoving sphere $r_{\rm AH}=1/aH$; spheres inside it are normal and spheres outside it anti-trapped~\cite{Bousso:2002ju}. During inflation with constant $\epsilon=-\dot H/H^2$ it shrinks as $r_{\rm AH}=-(1-\epsilon)\eta$, with $\eta\to0^-$ the future boundary of the inflating region. If inflation lasts $N=\ln(a_{\rm end}/a_{\rm i})$ e-folds, the modes that were trans-Planckian at the start of inflation have comoving wavenumbers $k\geq a_{\rm i}\Mp$ and those outside the horizon at the end have $k\leq a_{\rm end}H_{\rm end}$; the TCC demands that the two sets do not overlap~\cite{Bedroya:2019snp},
\begin{equation}\label{eq:classical-modes}
\{k:\,a_{\rm i}\Mp\leq k\leq a_{\rm end}H_{\rm end}\}=\varnothing\quad\Longleftrightarrow\quad e^{N}<\frac{\Mp}{H_{\rm end}}\,.
\end{equation}
Geometrically, the TCC holds if and only if inflation ends before the shrinking apparent horizon reaches the comoving sphere $r_{\rm P}=1/(a_{\rm i}\Mp)$ that was Planck-sized at the start, that is, before the scale factor has grown by $\Mp/H_{\rm end}$. The deadline is thus the fixed spacelike surface $a=a_{\rm i}\Mp/H_{\rm end}$ (Fig.~\ref{fig:penrose}(a)), and in e-folds it is Eq.~\eqref{eq:intro-tcc}.

When the end of inflation is stochastic the reheating surface is ragged, but the deadline stays a fixed surface: the one on which every comoving worldline has expanded by the factor $\Mp/H_{\rm end}$ since the start. A patch that inflates longer reheats closer to the future boundary $\eta=0$ (Fig.~\ref{fig:penrose}(b)--(d)), and since $H_{\rm end}$ is fixed by the end surface in field space, the deadline in e-folds is the same for every history: the TCC becomes the boundary condition of a first-passage problem.

The deadline is logarithmic in the entropy, while the bounds that the de Sitter entropy places on the number of e-folds and on the reheated volume of inflation that is not eternal~\cite{Arkani-Hamed:2007ryv,Dubovsky:2008rf,Dubovsky:2011uy} are linear in it. Up to an additive constant the deadline is also half the de Sitter scrambling time~\cite{Sekino:2008he,Bedroya:2020rmd,Aalsma:2020aib}, a motivation that recent results on anti-scrambling in de Sitter space call into question~\cite{Cui:2026bcd,Harlow:2026pwe}; we use only its logarithmic form. For any deadline affine in $\ln(\Mp/H_{\rm end})$ the power laws in $H_{\rm end}/\Mp$ found below acquire constant prefactors and exponents rescaled by a common factor, and their thresholds are unchanged; an example is $N<\ln S_{\rm end}$~\cite{Aalsma:2020aib}.

\subsection{Classical consequences}\label{sec:classical-consequences}
Three consequences will be revisited. First, the present Hubble radius left the horizon $N_+$ e-folds before the end of inflation, with $e^{N_+}\simeq V^{1/4}/(\mathcal T_0\mathcal T_{\rm eq})^{1/2}$ for rapid reheating in terms of the present and equality temperatures, and $e^{N_+}<\Mp/H$ gives~\cite{Bedroya:2019tba}
\begin{equation}\label{eq:classical-bblv}
V^{3/4}<\sqrt3\,\Mp^2(\mathcal T_0\mathcal T_{\rm eq})^{1/2}\,,
\end{equation}
the bounds on $V^{1/4}$ and $r$ quoted in the introduction, up to the values adopted for the temperatures ($V^{1/4}<5\times10^8\,$GeV for $\mathcal T_{\rm eq}\simeq0.8\,$eV).

Second, a de Sitter vacuum that decays by bubble nucleation at the rate $\Gamma$ per unit four-volume kills a comoving worldline at the rate $\kappa_{\rm d}=4\pi\Gamma/(3H^4)$ per e-fold~\cite{Guth:1982pn}. The false-vacuum volume grows, and inflation is eternal, if $\kappa_{\rm d}<3$, that is $\Gamma<9H^4/(4\pi)$. Read as a bound on the mean lifetime, $1/\kappa_{\rm d}\leq N_{\rm TCC}$, the TCC requires a decay rate only $3N_{\rm TCC}$ times smaller than this threshold. The stronger requirement $\Gamma\gtrsim H^4$ of Ref.~\cite{Bedroya:2020rac} drops the logarithmic factor of the lifetime bound at leading order; with it kept, the classical TCC does not forbid eternal false-vacuum inflation.

Third, rolling potentials dual to cascades of de Sitter bubbles have been argued, from the TCC, to obey $|V'|>C\,V^{3/2}$ with an undetermined constant $C$~\cite{Bedroya:2020rac}, the form of the condition for the absence of eternal inflation. Since $\PZ=V^3/(12\pi^2V'^2)$, a bound $\PZ<\mathcal P_{\rm c}$ is the slope bound with $C=(12\pi^2\mathcal P_{\rm c})^{-1/2}$, and the onset of eternal inflation on a linear slope, $\PZ=1/6$~\cite{Creminelli:2008es}, gives $C=1/(\sqrt2\,\pi)$~\cite{Rudelius:2019cfh}. The classical argument does not fix $C$, and its exclusion of eternal inflation is correspondingly only marginal.

\section{Stochastic Censorship}\label{sec:stoch}
\subsection{Durations and readings}\label{sec:stoch-readings}
In the overdamped limit the coarse-grained inflaton of a Hubble patch obeys the Langevin equation~\cite{Starobinsky:1986fx,Starobinsky:1994bd,Cruces:2022imf}
\begin{equation}\label{eq:stoch-langevin}
\dd\phi=-\frac{V'}{V}\,\dd N+\frac{H}{2\pi}\,\dd W_N\,,\qquad H^2=\frac V3\,,
\end{equation}
with $W_N$ a Wiener process in e-folds, drift $b=-V'/V$ and diffusion coefficient $D=H^2/(8\pi^2)$. The duration $\cN$ of a history is the time of its first passage through the end surface, and its survival function $\Pr[\cN>N]$ obeys the backward Fokker--Planck equation with absorption at the end surface~\cite{Vennin:2015hra,Pattison:2017mbe}. Expanded in the eigenfunctions of the generator of that equation, $\mathcal L\psi_n=-\Lambda_n\psi_n$, with coefficients $a_n$ (Appendix~\ref{app:worldline})~\cite{Ezquiaga:2019ftu}, it gives at the deadline
\begin{equation}\label{eq:stoch-spectral}
\Pr[\cN>N_{\rm TCC}]=\sum_{n}a_n\psi_n(\phi_*)\Big(\frac{H_{\rm end}}{\Mp}\Big)^{\Lambda_n}\,,
\end{equation}
for a history that starts at $\phi_*$; on a half-line the sum becomes an integral over a continuum above $\Lambda_0$. Because the deadline is logarithmic, exponential tails of the duration become powers of $H_{\rm end}/\Mp$, with the spectral gap $\Lambda_0$ as exponent. For a linear roll (constant drift and diffusion), which we also call a linear slope, the duration has the inverse-Gaussian law, with mean $N_{\rm cl}$, variance $\PZ N_{\rm cl}$ and gap
\begin{equation}\label{eq:stoch-gap}
\Lambda_0=\frac{b^2}{4D}=\frac1{2\PZ}\,.
\end{equation}

A probabilistic TCC needs three choices. The first is a confidence level. Every deadline is exceeded with positive probability, so a TCC imposed on all histories would exclude every stochastic theory, including those whose classical histories satisfy it by tens of e-folds. We ask instead that a violation have probability at most $\delta$, held fixed as $H_{\rm end}/\Mp$ varies, as in Ref.~\cite{Anchordoqui:2026edd}; a probabilistic TCC extends the conjecture to a quantum spread of histories and is not a consequence of it. The second is the initial condition: a Hubble patch at a given field value, from which the classical history lasts $N_{\rm cl}$ e-folds. The third is the reading. A patch that inflates for $N$ e-folds becomes $e^{3N}$ patches, and $e^{(3-\kappa_{\rm d})N}$ on average if the vacuum decays at the rate $\kappa_{\rm d}$ per e-fold. We compare
\begin{itemize}
\item the \emph{worldline} reading, $P^{\rm wl}=\Pr[\cN>N_{\rm TCC},\ \text{no decay}]$: is the Hubble patch of one
comoving worldline still inflating at the deadline?
\item the \emph{region} reading, $P^{\rm reg}=\Pr[Z_{N_{\rm TCC}}>0]$, with $Z_N$ the number of descendants of the initial patch that still inflate at $N$: does any trans-Planckian mode of the region grown from that patch cross the horizon after the deadline?
\item the \emph{volume} reading, the fraction of the reheated volume whose history ends after the deadline,
\begin{equation}\label{eq:stoch-vol}
P^{\rm vol}=\frac{\mathbb E\big[e^{(3-\kappa_{\rm d})\cN}\,\mathbf 1_{\cN>N_{\rm TCC}}\big]}{\mathbb E\big[e^{(3-\kappa_{\rm d})\cN}\big]}\,,
\end{equation}
by the many-to-one lemma of branching diffusions~\cite{hardy2006newformulationspineapproach,10.1214/15-AIHP714}. Patches removed by decay do not reheat, except in a false vacuum without roll, where decay is the only exit and the volume that enters bubbles is
counted instead (Section~\ref{sec:metastable}). The normalization diverges when inflation is eternal, and the reading then needs a regulator.
\end{itemize}
For the region we use, in the separate-universe approximation, a branching diffusion, in the spirit of tree models of eternal inflation~\cite{Harlow:2011az}: a patch splits at the rate $3$ into patches with independent later noise, which neglects correlations below the Hubble radius at the splitting (Appendix~\ref{app:regions}). Every reading reduces to the classical TCC when the noise vanishes. The first two are ordered by elementary facts, which hold in any model: a line of descent chosen at random is a worldline, and $\mathbf 1_{Z>0}\leq Z$, so that
\begin{equation}\label{eq:stoch-union}
P^{\rm wl}\;\leq\;P^{\rm reg}\;\leq\;\mathbb E[Z_{N_{\rm TCC}}]=e^{3N_{\rm TCC}}\,P^{\rm wl}\,.
\end{equation}
The factor $e^{3N_{\rm TCC}}=(\Mp/H_{\rm end})^3$ counts the Hubble patches at the deadline, each of which descends from a cell that was Planck-sized at the start: the region reading at confidence $1-\delta$ is implied by the worldline reading at confidence $1-\delta\,(H_{\rm end}/\Mp)^3$, as if the worldline condition were imposed on every such cell. We impose the TCC at the deadline. Along a classical slow roll earlier times give weaker conditions; imposing it at all earlier times tightens the deadline by at most $\ln(H_{\max}/H_{\rm end})$ e-folds when the field makes upward excursions, with $H_{\max}$ the largest Hubble rate reached, and not at all on a linear slope.

\subsection{Which reading?}\label{sec:stoch-which}
The conjecture as stated, that no trans-Planckian mode ever crosses the horizon, refers to all modes of the expanding region; its literal translation is the region reading. Its motivations from de Sitter complementarity and from the scrambling time of the horizon~\cite{Bedroya:2020rmd,Aalsma:2020aib} refer instead to the patch of one observer: the worldline reading, or an intermediate one for an observer who sees the region grown from one Hubble patch at a later time $N_0$. That region has $e^{3(N_{\rm TCC}-N_0)}$ patches at the deadline on average, which replaces $w=3$ by $w=3(1-N_0/N_{\rm TCC})$ in Eq.~\eqref{eq:stoch-main} below. For our observable universe, which left the horizon $N_+$ e-folds before the end of a classical history that nearly saturates the deadline, $w\simeq3N_+/N_{\rm TCC}$, close to the region value when the bound \eqref{eq:classical-bblv} on the scale of inflation is nearly saturated; an initial region of $n$ Hubble patches has $w=3+\ln n/N_{\rm TCC}$. The volume reading is a measure on the reheated volume, with the local and global choices familiar from eternal inflation~\cite{Bousso:2006ev,Garriga:2005av,Bousso:2009mw}, and its normalization involves the evolution after the deadline. We do not choose between the readings and state the results reading by reading. The worldline reading asks that the observer on one comoving worldline see no violation, the region reading that no observer in the region see one; swampland conditions, stated for every consistent history, are closest to the latter. Part of the readings' verdicts on eternal inflation is built into the definitions, since one worldline leaves an eternal regime almost surely while a region that inflates forever with probability above $\delta$ violates the TCC by construction; what the analysis adds is where the thresholds lie and how they depend on the classical history. Patches that inflate past the deadline while their surroundings reheat carry a curvature perturbation of at least $N_{\rm TCC}-N_{\rm cl}$ and may end inside black holes or baby universes~\cite{Tomberg:2025fku,Garriga:2015fdk}; the region reading counts them, and a single-observer reading does not.

\subsection{The stochastic TCC as an inequality}\label{sec:stoch-ineq}
For a linear roll the moment generating function of the duration is finite up to $s=\Lambda_0$, and Markov's inequality applied to $e^{s\cN}$, optimized over $s$, gives for every $T\geq N_{\rm cl}$ (Appendix~\ref{app:worldline-chernoff})
\begin{equation}\label{eq:stoch-chernoff}
\Pr\big[\cN>T\big]\;\leq\;\exp\Big[-\frac{(T-N_{\rm cl})^2}{2\PZ\,T}\Big]\,.
\end{equation}
At $T=N_{\rm TCC}$ this bounds the worldline reading, and through Eq.~\eqref{eq:stoch-union} the region reading. In terms of the fraction $c\equiv N_{\rm cl}/N_{\rm TCC}$ of the deadline used by the classical history, the TCC therefore holds at confidence $1-\delta$ whenever
\begin{equation}\label{eq:stoch-main}
\boxed{\;1-c\;\geq\;\sqrt{2\PZ\Big(w+\frac{\ln(1/\delta)}{N_{\rm TCC}}\Big)}\;}\,,
\qquad w=\begin{cases}0&\text{worldline}\,,\\3&\text{region}\,.\end{cases}
\end{equation}
The left-hand side is the unused fraction of the deadline, which the classical TCC only asks to be positive. Per worldline the stochastic TCC asks for the margin $[2\PZ N_{\rm TCC}\ln(1/\delta)]^{1/2}$, a few standard deviations of the duration and a vanishing fraction of a long deadline; for a region it asks for at least the fraction $\sqrt{6\PZ}$, whatever the deadline, which at leading order is also necessary. A decay rate $\kappa_{\rm d}$ replaces $w$ by $w-\kappa_{\rm d}$, and when $w-\kappa_{\rm d}+\ln(1/\delta)/N_{\rm TCC}$ is then negative the condition holds for every $c$. The confidence enters only through $\ln(1/\delta)/N_{\rm TCC}$: at fixed $\delta$ its effect is logarithmic and disappears at leading order, while a confidence that tightens with the deadline, $\delta=(H_{\rm end}/\Mp)^p$, shifts $w$ by $p$. The region condition is the worldline condition tightened in this way with $p=3$, through Eq.~\eqref{eq:stoch-union}.

The volume reading has the same structure. Weighting by $e^{(3-\kappa_{\rm d})\cN}$ maps the inverse-Gaussian family onto itself: the volume-weighted duration is that of a slower roll, with mean $N_{\rm cl}/\sqrt{1-\chi}$ and power spectrum
$\PZ/(1-\chi)$, where
\begin{equation}\label{eq:stoch-chi}
\chi\equiv2(3-\kappa_{\rm d})\PZ=\frac{3-\kappa_{\rm d}}{\Lambda_0}
\end{equation}
is at most one exactly when inflation is not eternal~\cite{Creminelli:2008es,Winitzki:2008ph}. Equation~\eqref{eq:stoch-chernoff} for the slower roll gives
\begin{equation}\label{eq:stoch-volume}
\sqrt{1-\chi}-c\;\geq\;\sqrt{\frac{2\PZ\ln(1/\delta)}{N_{\rm TCC}}}\,,
\end{equation}
at leading order the worldline condition with the deadline shortened by the factor $\sqrt{1-\chi}$. For $\chi>1$ the volume reading needs a regulator (Section~\ref{sec:slope-regulators}).

For a slowly varying power spectrum the worldline condition takes a Bernstein form. At leading order in the slow variation the cumulant generating function of the duration adds up linear-roll contributions along the classical path. Each exceeds $s\,\dd N$, its value without noise, by $\PZ$ times a weight that grows with $\PZ$, and bounding the weight by its value at the largest power spectrum $\mathcal P_{\max}$ gives (Appendix~\ref{app:worldline-chernoff})
\begin{equation}\label{eq:stoch-bernstein}
\Pr\big[\cN>N_{\rm cl}+t\big]\;\lesssim\;\exp\Big[-\frac{t^2}{2(\sigma_N^2+\mathcal P_{\max}t)}\Big]\,,\qquad
\sigma_N^2=\int_0^{N_{\rm cl}}\PZ\,\dd N\,,
\end{equation}
with $\sigma_N^2$ the variance of the duration in the $\delta N$ formalism and $\lesssim$ marking the adiabatic approximation. The worldline condition is then
\begin{equation}\label{eq:stoch-general}
N_{\rm TCC}-N_{\rm cl}\;\gtrsim\;\mathcal P_{\max}\ln\frac1\delta+\Big[\mathcal P_{\max}^2\ln^2\frac1\delta+2\sigma_N^2\ln\frac1\delta\Big]^{1/2}\,,
\end{equation}
which reduces to Eq.~\eqref{eq:stoch-main} with $w=0$ for a constant power spectrum. The term in $\mathcal P_{\max}$ is needed: the Gaussian estimate alone underestimates the tail, while the bound holds against direct simulations for margins large compared with $\mathcal P_{\max}$. A region probes a much deeper tail, which is set by the spectral gap of the whole field range it can explore, including excursions up the potential; $\mathcal P_{\max}$ must then be taken over that range.

\subsection{Sharpness and the order of the readings}\label{sec:stoch-sharp}
Let $N_{\rm TCC}\to\infty$ at fixed $c$, $\delta$ and $\kappa_{\rm d}<3$. To leading order the conditions become
\begin{equation}\label{eq:stoch-leading}
\text{classical, worldline: } c<1\,,\qquad\text{volume: } c<\sqrt{1-\chi}\,,\qquad
\text{region: } c\leq1-\sqrt\chi\,,
\end{equation}
with $\chi=6\PZ$ without decay; with decay the worldline reading imposes no condition at this order, since decay alone ends the worldline in time. These conditions are also necessary (for the region reading with decay, as long as $1-\kappa_{\rm d}/3>\delta$), so the inequality is sharp at leading order in every reading. Per worldline, the Chernoff exponent is the exact large deviation of the inverse Gaussian, and only a prefactor of order $N_{\rm TCC}^{-1/2}$ is lost; near $c=1$ the exact margin is close to $\Phi^{-1}(1-\delta)$ standard deviations rather than $[2\ln(1/\delta)]^{1/2}$, with $\Phi$ the standard normal distribution function, a sixth
fewer at $\delta=10^{-3}$. For the volume, the tilted law is exact. For a region, the sharpness is a theorem (Appendix~\ref{app:regions-front}): for binary splitting,
\begin{equation}\label{eq:stoch-step}
\lim_{N_{\rm TCC}\to\infty}P^{\rm reg}=\begin{cases}1-\kappa_{\rm d}/3\,,&c>1-\sqrt\chi\,,\\0\,,&c\leq1-\sqrt\chi\,,\end{cases}
\end{equation}
where $1-\kappa_{\rm d}/3$ is the probability that decay does not remove all patches. Below the threshold, the bound $\mathbb E[Z]$ vanishes; above it the region is compared with a population absorbed on a line that moves more slowly than the front of branching Brownian motion~\cite{Mckean1975ApplicationOB,VANSAARLOOS_2003}, and therefore survives. Figure~\ref{fig:bounds}(b) shows the approach to the step, from the equation for $P^{\rm reg}$ in Appendix~\ref{app:regions-fkpp}. At finite deadlines, the half-height of the step lies at larger $c$, by an amount that falls as $\ln N_{\rm TCC}/N_{\rm TCC}$, the logarithmic delay of the front~\cite{VANSAARLOOS_2003}, while the level $P^{\rm reg}=\delta$ stays close to $1-\sqrt\chi$. The inequality gives away about one e-fold or less at the deadlines shown: at $N_{\rm TCC}=11.5$ it allows $c\leq0.24$ where the exact region condition allows $c\leq0.30$.

Since $1-\sqrt\chi\leq\sqrt{1-\chi}\leq1$, the region reading is, at leading order, the strictest and the worldline reading the weakest for a classical history that uses a finite fraction of the deadline. The order of region and volume is not universal: for a
false vacuum, where $P^{\rm vol}$ equals the expected count $\mathbb E[Z]\geq P^{\rm reg}$, and for a patch near the end surface of a slope that is not eternal, the volume reading is the stricter one.

Is the stochastic TCC ever weaker than the classical one? On a linear slope without decay the worldline and region readings never are, at least for $\PZ<2\pi N_{\rm TCC}(1-2\delta)^2$, far beyond the eternal onset. A classical history that ends exactly at the deadline violates the worldline reading with probability close to one half, at least $\frac12-[\PZ/(8\pi N_{\rm TCC})]^{1/2}$ by Eq.~\eqref{eq:app-ig} and the Mills-ratio bound, and longer histories violate it more often. For any $\delta$ below this value the worldline reading therefore implies $N_{\rm cl}<N_{\rm TCC}$; so does the region reading, since $P^{\rm reg}\geq P^{\rm wl}$, and so does the volume reading wherever its normalization is finite, since the volume weight favors long histories. The same holds at leading order in the slow variation on any slope on which the duration is close to normal. The stochastic TCC is weaker in three cases, each with a caveat. In a flat quantum well the overdamped classical duration diverges and the classical TCC excludes the model outright, while the worldline reading, and for a well below the eternal size also the region reading, impose finite margins (Section~\ref{sec:wells}); a slight tilt makes the classical duration long rather than infinite and leaves the margins nearly unchanged. With decay a patch can stop inflating before its classical roll ends. Counting a worldline that decays as one whose inflation has ended, the worldline reading accepts histories longer than the deadline once $\kappa_{\rm d}N_{\rm TCC}>\ln(1/\delta)$, which relaxes the roll condition $N_{\rm cl}<N_{\rm TCC}$ but is stricter, by the factor $\ln(1/\delta)$, than the mean-lifetime reading of Section~\ref{sec:classical-consequences}; conditioned instead on reheating by the roll, which tilts the distribution of the duration by $e^{-\kappa_{\rm d}\cN}$, the worldline reading relaxes the leading-order condition only to $c<(1+2\kappa_{\rm d}\PZ)^{1/2}$. Where inflation is eternal, the volume reading with the finite-history regulator can accept models that the classical TCC forbids, but only on the event that the region stops inflating, which can be rarer than the violations it tolerates (Section~\ref{sec:slope-regulators}).

These statements hold within leading-order, overdamped stochastic inflation and, for a region, the branching model; ``exact'' here and throughout means exact within it. The region reading probes the duration far in its tail: by Eq.~\eqref{eq:stoch-union} it involves per-worldline probabilities of order $\delta\,(H_{\rm end}/\Mp)^3$, about $10^{-63}$ at $N_{\rm TCC}=46$. Such a deviation is reached by a sustained bias of the noise against the drift, about $\sqrt6$ standard deviations per e-fold at the threshold, not by extreme single kicks, so for $\PZ\ll1$ it stays within the Gaussian regime of the coarse-grained noise. Near the onset of eternal inflation, where $\PZ$ is of order $0.1$, a correction of relative size $\tau\PZ$ to the gap, which the leading-order description neglects, would move the region threshold by about $\tau\sqrt6\,\PZ^{3/2}/2$ of the deadline, some $2\tau$ e-folds at $\PZ=0.1$ and $N_{\rm TCC}=46$, comparable to or larger than the difference between the sufficient and the exact conditions.

Figure~\ref{fig:bounds}(a) shows the unused fraction $1-c$ that suffices, and as $N_{\rm TCC}\to\infty$ is also necessary; at $\PZ=1/6$ the region and volume readings require the whole deadline to remain unused. Figure~\ref{fig:penrose}(b)--(d) shows the regimes in one realization.

\begin{figure}[htb]\centering
\includegraphics[width=\textwidth]{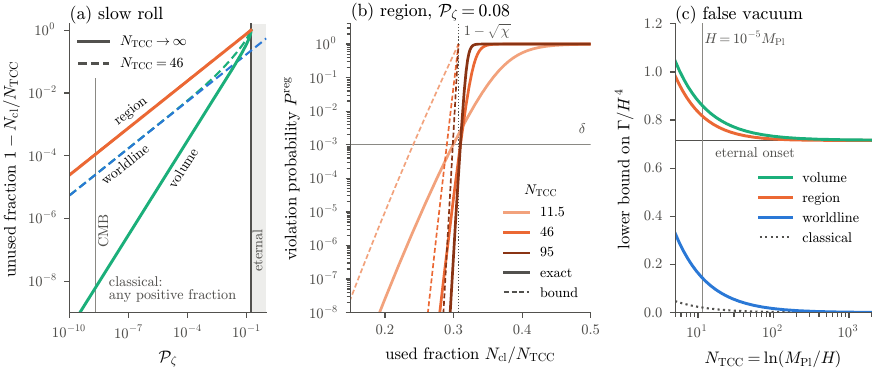}
\caption{The stochastic TCC at confidence $1-\delta$, $\delta=10^{-3}$.
(a) Slow roll: the unused fraction of the deadline, $1-N_{\rm cl}/N_{\rm TCC}$, that suffices for the TCC in the region,
volume and worldline readings, from Eqs.~\eqref{eq:stoch-main} and \eqref{eq:stoch-volume} at $H_{\rm end}=10^{-20}\Mp$,
$N_{\rm TCC}=46$ (dashed), and the fraction that is necessary and sufficient as $N_{\rm TCC}\to\infty$, Eq.~\eqref{eq:stoch-leading} (solid), where the worldline fraction vanishes and is not drawn. At this deadline, the two region curves nearly coincide, and at small $\PZ$ the dashed volume curve follows the dashed worldline curve. The thin vertical line marks the CMB normalization. Beyond the onset of eternal inflation, $\PZ>1/6$ (shaded), only the worldline condition continues (Section~\ref{sec:slope-eternal}). (b) Region reading at $\PZ=0.08$: the violation probability against the fraction of the deadline used, at three deadlines, exact from Eq.~\eqref{eq:app-fkpp} (solid) and its bound $e^{3N_{\rm TCC}}P^{\rm wl}$ with Eq.~\eqref{eq:stoch-chernoff} (dashed), which crosses $\delta$ where Eq.~\eqref{eq:stoch-main} with $w=3$ is saturated. Both approach the step \eqref{eq:stoch-step} at $1-\sqrt\chi$ (dotted). (c) False vacuum: lower bounds on the decay rate, Eqs.~\eqref{eq:metastable-thresholds} and \eqref{eq:metastable-region}, and the classical
mean-lifetime bound (dotted), against $N_{\rm TCC}=\ln(\Mp/H)$; the horizontal line is the eternal onset, and the thin vertical line marks $H=10^{-5}\Mp$ of Table~\ref{tab:metastable}.}
\label{fig:bounds}
\end{figure}

\section{Slow Roll}\label{sec:slope}
\subsection{The attractor and the scale of inflation}\label{sec:slope-cmb}
On the attractor normalized to the CMB, $\PZ\simeq2.1\times10^{-9}$, the noise is a small correction to the drift and the inverse Gaussian is nearly normal. The worldline condition is then, to excellent accuracy,
\begin{equation}\label{eq:slope-margin}
N_{\rm TCC}-N_{\rm cl}\;\geq\;\Phi^{-1}(1-\delta)\,\sqrt{\PZ N_{\rm cl}}\,,
\end{equation}
a thousandth of an e-fold for a deadline of a few tens of e-folds and $\delta=10^{-3}$. The region margin $\sqrt{6\PZ}\,N_{\rm TCC}$, the extreme-value spread of the durations of the $e^{3N_{\rm TCC}}$ patches of the region, is about a hundredth of an e-fold or less. In the argument behind Eq.~\eqref{eq:classical-bblv} a margin $\Delta$ multiplies the bound on $V^{3/4}$ by $e^{-\Delta}$, so the bounds on the scale of inflation and on $r$ hold in every reading, with corrections below a percent, far smaller than their dependence on the assumed history around inflation~\cite{Mizuno:2019bxy,Kamali:2019gzr,Brandenberger:2019eni}. Here, as throughout the paper, observables follow from the stochastic $\delta N$ formalism~\cite{Fujita:2013cna,Vennin:2015hra}, which with subdominant noise reproduces the standard $\PZ=H^2/(8\pi^2\epsilon)$, $n_s$ and $r$ up to relative corrections of order $\PZ$; noise matters for observables only where $\PZ$ is not small, notably through exponential tails of the distribution of $\zeta$~\cite{Pattison:2017mbe,Ezquiaga:2019ftu}. The TCC, by contrast, constrains the whole duration, including its earliest, unobservable e-folds and its tail.

\subsection{Eternal inflation and the slope bound}\label{sec:slope-eternal}
Near the onset of eternal inflation the readings differ. For a region, Eq.~\eqref{eq:stoch-main} can be satisfied by some $c\geq0$ only if $\PZ\leq1/[6+2\ln(1/\delta)/N_{\rm TCC}]$, below the onset; beyond this value the inequality certifies no history. For $\PZ\geq1/6$ the region reading indeed fails, by the step \eqref{eq:stoch-step}: its threshold $1-\sqrt\chi$ is not positive, so a classical history that uses any finite fraction of the deadline violates the region reading with a probability that tends to one. The region reading of the TCC therefore excludes eternal inflation on a linear slope for every classical history that uses a finite fraction of the deadline. For a varying power spectrum the same holds for the field range that the region can explore: if it contains
an eternal stretch, the violation probability is at least the probability that some descendant reaches the stretch and
inflates there forever, which does not depend on the deadline.

Solved for the slope, the region and worldline conditions on a linear slope are bounds of the classical form $|V'|>C\,V^{3/2}$, now with a
definite coefficient,
\begin{equation}\label{eq:slope-C}
C=\frac{1}{\pi(1-c)}\sqrt{\frac{w+\ln(1/\delta)/N_{\rm TCC}}{6}}\;\xrightarrow{\;N_{\rm TCC}\to\infty\;}\;
\begin{cases}\dfrac{1}{\sqrt2\,\pi\,(1-c)}&\text{region}\,,\\[8pt]0&\text{worldline}\,,\end{cases}
\end{equation}
sufficient at every $N_{\rm TCC}$ and sharp at leading order; the volume reading gives $1/[\sqrt2\,\pi(1-c^2)^{1/2}]$ at leading order. For a classical history that uses a vanishing fraction of the deadline the region coefficient is the onset value~\cite{Rudelius:2019cfh}, and the region reading reproduces the known condition for the absence of eternal inflation. What
the TCC adds is the factor $(1-c)^{-1}$: the coefficient depends on the history through the fraction of the deadline it uses, not on the potential at a point. For a roll across a field range $\Delta\phi$ the fraction used is $c=\Delta\phi\,V/(|V'|N_{\rm TCC})$, and at leading order the region condition becomes $|V'|>V^{3/2}/(\sqrt2\,\pi)+\Delta\phi\,V/N_{\rm TCC}$: the bounds from the absence of eternal inflation and from the classical TCC add up, and in the volume reading they add in quadrature. A classical history that uses the fraction $c$ violates the region reading, with a probability that tends to one, for every slope in the band
\begin{equation}\label{eq:slope-band}
\frac{(1-c)^2}{6}\;<\;\PZ\;<\;\frac16\,,
\end{equation}
so the TCC excludes non-eternal slopes next to the onset. Per worldline, in contrast, every slope is allowed for classical histories that end a few standard deviations of the duration before the deadline.

The eternal side has a structure of its own when the patch starts a fixed number of e-folds from the end surface. Then $P^{\rm reg}$ decreases toward the probability $P_{\rm et}$ that some descendant inflates forever, and
\begin{equation}\label{eq:slope-pet}
P_{\rm et}>0\quad\Longleftrightarrow\quad\Lambda_0+\kappa_{\rm d}<3\,,
\end{equation}
the spectral form of the onset~\cite{Winitzki:2001np,Winitzki:2008ph} and, for branching Brownian motion with absorption,
a theorem~\cite{KESTEN19789}; Appendix~\ref{app:regions-pet} gives a short proof for any slope on a bounded domain, away from the critical case. The equation for $P_{\rm et}$ shows that it is of order one beyond a correlation length that diverges at the onset and
exponentially small closer to the end surface (Appendix~\ref{app:regions-fkpp},~\cite{Berestycki_2011}). The region reading therefore excludes the eternal regimes with $P_{\rm et}>\delta$ and allows, at long enough deadlines, patches that start so close to the end surface that $P_{\rm et}<\delta$. A classical history that uses a finite fraction of the deadline is excluded in every eternal regime, since then $P^{\rm reg}\to1-\kappa_{\rm d}/3$, unless decay alone nearly ends inflation, $1-\kappa_{\rm d}/3\leq\delta$ (Section~\ref{sec:metastable}).

At the threshold $\Lambda_0+\kappa_{\rm d}=3$ the violation probability decays more slowly than any power of $H_{\rm end}/\Mp$,
\begin{equation}\label{eq:slope-kesten}
P^{\rm reg}\simeq A\,X e^{QX/2}\exp\Big[-\big(3\pi^2\Lambda_0N_{\rm TCC}\big)^{1/3}\Big]\,,
\end{equation}
as $N_{\rm TCC}\to\infty$ at fixed distance $X=3QN_{\rm cl}$ from the end surface, in the units of Appendix~\ref{app:regions}, where $Q=[2/(3\PZ)]^{1/2}$, with $A$ a constant set by the branching law. The stretched exponential was established in Ref.~\cite{KESTEN19789} up to corrections of lower order in the exponent; its prefactor was derived heuristically in Ref.~\cite{Derrida_2007}, proved up to constants in Ref.~\cite{berestycki2014critical}, and proved as an asymptotic equality in Ref.~\cite{AHL_2022__5__921_0} for supercritical offspring laws with finite variance. Numerical solutions of Eq.~\eqref{eq:app-fkpp} at the threshold reproduce both the exponent and the profile $Xe^{QX/2}$, with and without decay (Appendix~\ref{app:lattice}). On the non-eternal side, for binary splitting without decay and at fixed distance from the end surface, $P^{\rm reg}$ is asymptotically proportional to $\mathbb E[Z_{N_{\rm TCC}}]$, as shown in Ref.~\cite{10.1214/ECP.v12-1259}: the upper bound of Eq.~\eqref{eq:stoch-union} is attained up to a constant.

\subsection{Volume regulators}\label{sec:slope-regulators}
On the non-eternal side the volume reading is the slower roll of Section~\ref{sec:stoch-ineq}, and the scale-factor cutoff and the finite-history regulator described below agree with it. On the eternal side its normalization diverges, and the answer depends on the regulator. The scale-factor cutoff~\cite{DeSimone:2008bq,DeSimone:2008if,Bousso:2008hz} counts the volume that reheats before a
cutoff time and then removes the cutoff; since the tilted density of durations grows exponentially, the violating fraction tends to one; a cutoff placed at the deadline itself would remove the violation by fiat. The reheating-volume cutoff~\cite{Winitzki:2008yb,Winitzki:2008ph} conditions on the reheated volume, $\mathcal V\to\infty$; histories shorter than the deadline reheat at most $e^{3N_{\rm TCC}}$ units, so its violating fraction tends to one in every regime, and no model satisfies the TCC. (The breakdown of volume renormalization at the onset of eternal inflation has recently been related to the central-charge-one barrier of Liouville gravity~\cite{Cotler:2026lna}.) The finite-history regulator, which weights histories by $e^{-z\mathcal V}$ and lets $z\to0^+$, keeps only histories that end. Among them the violation is suppressed as $(H_{\rm end}/\Mp)^{\Lambda_{\rm fin}}$ with $\Lambda_{\rm fin}>0$ for every slope, because the conditioning removes exactly the eternal histories (Appendix~\ref{app:regions-fin}). The exponent vanishes at the onset as $(6/\pi)[1-1/(6\PZ)]^{3/2}$ at leading order and is $4.61$ on a flat potential. This regulator also shortens the histories it keeps: on a flat potential, which the classical TCC excludes outright because the classical field never moves, the TCC in this reading is satisfied once $N_{\rm TCC}\gtrsim\ln(1/\delta)/\Lambda_{\rm fin}$, and in this sense it is weaker than the classical TCC. The acceptance holds on the event that the region stops inflating, whose probability in the model of Appendix~\ref{app:regions-fin}, $1-P_{\rm et}=e^{-q_\infty}$, falls on a flat potential below $\delta=10^{-3}$ for a patch more than about $H/4$ from the end surface; there the reading accepts the model on an event rarer than the violations it tolerates. The volume reading of eternal inflation is thus a statement about the measure, while the region reading needs no
regulator.

\section{Metastable de Sitter Space}\label{sec:metastable}
A false vacuum with Hubble rate $H$ that decays at the rate $\kappa_{\rm d}=4\pi\Gamma/(3H^4)$ per patch per e-fold is the limit without roll: a worldline inflates until a bubble nucleates in its past light cone, and the deadline is $N_{\rm TCC}=\ln(\Mp/H)$. We let patches die independently, which neglects the growth of a bubble into neighboring patches; within this model the readings are exact for every $N_{\rm TCC}$. The lifetime of a worldline is exponential, and so is its tilt by $e^{3N}$; counting the volume that enters bubbles,
\begin{equation}\label{eq:metastable-readings}
P^{\rm wl}=\Big(\frac{H}{\Mp}\Big)^{\kappa_{\rm d}}\,,\qquad
\mathbb E[Z_{N_{\rm TCC}}]=\Big(\frac{\Mp}{H}\Big)^{3-\kappa_{\rm d}}\,,\qquad
P^{\rm vol}=\Big(\frac{H}{\Mp}\Big)^{\kappa_{\rm d}-3}\quad(\kappa_{\rm d}>3)\,.
\end{equation}
For patches that split in two at the rate $3$ the survival probability $u(N)=\Pr[Z_N>0]$ obeys the logistic equation $\dd u/\dd N=(3-\kappa_{\rm d})u-3u^2$, with $u(0)=1$, so that
\begin{equation}\label{eq:metastable-reg}
P^{\rm reg}=\frac{3-\kappa_{\rm d}}{3-\kappa_{\rm d}\,e^{-(3-\kappa_{\rm d})N_{\rm TCC}}}\,,
\end{equation}
which tends to $P_{\rm et}=1-\kappa_{\rm d}/3$ on the eternal side and equals $1/(1+3N_{\rm TCC})$ at the onset.

Each reading turns the TCC into a lower bound on the decay rate. The worldline and volume bounds are
\begin{equation}\label{eq:metastable-thresholds}
\frac{\Gamma_{\rm wl}}{H^4}=\frac{3}{4\pi}\,\frac{\ln(1/\delta)}{N_{\rm TCC}}\,,\qquad
\frac{\Gamma_{\rm vol}}{H^4}=\frac{9}{4\pi}+\frac{\Gamma_{\rm wl}}{H^4}\,,
\end{equation}
the second being the eternal onset shifted by the first, since $P^{\rm vol}=e^{3N_{\rm TCC}}P^{\rm wl}$. The region bound of binary splitting is $\Gamma_{\rm reg}/H^4=3(3+\xi)/(4\pi)$, with $\xi$ the nonzero root of
\begin{equation}\label{eq:metastable-region}
(3+\xi)\,e^{\xi N_{\rm TCC}}=3+\frac\xi\delta\,,\qquad
\xi\simeq\frac{1}{N_{\rm TCC}}\ln\Big[\frac{1}{3N_{\rm TCC}\delta}\ln\frac{1}{3N_{\rm TCC}\delta}\Big]\,,
\end{equation}
the approximation holding for $3N_{\rm TCC}\delta\ll1$. For any branching law with the same mean growth, the region bound lies below the volume bound, since $P^{\rm reg}\leq\mathbb E[Z]$, and less variable laws, in the convex order, survive more often and give higher bounds. A Poisson number of patches per e-fold gives a slightly higher bound than binary splitting. For the physical process, deterministic growth with nucleation, the law depends on how nucleations in neighboring patches are correlated; simple discretizations with the same mean growth give bounds between $0.78$ and $0.85$ at the parameters of Table~\ref{tab:metastable}, all below the volume bound.

\begin{table}[t]\centering\small
\begin{tabular}{lcc}\toprule
reading & condition & $\Gamma/H^4$\\\midrule
classical, mean lifetime & $1/\kappa_{\rm d}\leq N_{\rm TCC}$ & $0.021$\\
worldline & $e^{-\kappa_{\rm d}N_{\rm TCC}}\leq\delta$ & $0.143$\\
eternal onset (reference) & $\kappa_{\rm d}=3$ & $9/(4\pi)=0.716$\\
region, binary splitting & Eq.~\eqref{eq:metastable-region} & $0.816$\\
region, Poisson law & generating function iterated per e-fold & $0.842$--$0.855$\\
volume & $e^{(3-\kappa_{\rm d})N_{\rm TCC}}\leq\delta$ & $0.859$\\\bottomrule
\end{tabular}
\caption{Lower bounds on the decay rate of a false vacuum at $H=10^{-5}\Mp$ ($N_{\rm TCC}=11.5$) and $\delta=10^{-3}$, a high-scale vacuum at which the readings differ most. Figure~\ref{fig:bounds}(c) shows how the bounds approach the eternal onset at longer deadlines, and row (d) of Table~\ref{tab:examples} gives them at $N_{\rm TCC}=46$. The Poisson range comes from iterating over $11$ and $12$ e-folds.}
\label{tab:metastable}
\end{table}

Table~\ref{tab:metastable} and Fig.~\ref{fig:bounds}(c) collect the bounds. The classical bound with its logarithm kept and the worldline bound lie below the onset: per worldline the TCC allows eternal false-vacuum inflation, since the Hubble patch of any given observer decays in time with high probability while the false-vacuum volume grows forever. The volume bound lies above the onset at every deadline, so the volume reading forbids eternal false-vacuum inflation; its normalization then diverges, and the scale-factor cutoff gives a violating fraction that tends to one. The region reading forbids it as well for every vacuum with $P_{\rm et}>\delta$ and, for binary splitting, for every eternal vacuum as long as $N_{\rm TCC}<(1/\delta-1)/3$, where $\xi>0$ and the region bound lies above the onset. Beyond that deadline ($333$ for $\delta=10^{-3}$, far beyond the deadline of the present vacuum, about $140$), the region bound falls below the onset, toward $9(1-\delta)/(4\pi)$ where $P_{\rm et}=\delta$, and allows the narrow window
$3(1-\delta)<\kappa_{\rm d}<3$, because at the onset $P^{\rm reg}$ decays only as the inverse deadline.

A slope and a decay rate combine. The onset of eternal inflation is $\Lambda_0+\kappa_{\rm d}=3$, so slow-roll and false-vacuum eternal inflation are the two ends of one line, and Eq.~\eqref{eq:stoch-main} with $w=3-\kappa_{\rm d}$ gives, for $c\leq1$, a sufficient region condition that is sharp at leading order,
\begin{equation}\label{eq:metastable-plane}
\frac{\Gamma}{H^4}\;\geq\;\frac{3}{4\pi}\Big[3-\Lambda_0(1-c)^2+\frac{\ln(1/\delta)}{N_{\rm TCC}}\Big]\,,
\end{equation}
in which roll and decay add up: a steeper roll, or a classical history that ends earlier, relaxes the bound on the decay rate. At the onset without roll the critical law $1/(1+3N_{\rm TCC})$ of Eq.~\eqref{eq:metastable-reg} takes the place of the stretched exponential \eqref{eq:slope-kesten}; both decay more slowly than any power of $H/\Mp$.

\section{Quantum Wells and Primordial Black Holes}\label{sec:wells}
Where the drift is small, the gap is set by the size of the flat part of the potential, and the stochastic TCC departs from the classical one even per worldline. For a flat well, $V=V_0$ over a field range $\Delta\phi$ above the end surface with a reflecting
top, the survival function from the top is~\cite{Pattison:2017mbe}
\begin{equation}\label{eq:wells-flat}
\mathcal S(N)=\sum_{n\geq0}\frac{4(-1)^n}{(2n+1)\pi}\,e^{-(2n+1)^2\pi^2N/4\mu^2}\,,\qquad
\Lambda_0=\frac{\pi^2}{4\mu^2}\,,\qquad\mu^2=\frac{24\pi^2\Delta\phi^2}{V_0}\,,
\end{equation}
an alternating series bounded by its first term; in this section $\Lambda_0$ is the gap of the well. The well is eternal when $\Lambda_0<3$, that is $\mu^2>\pi^2/12$~\cite{Animali:2024jiz}, the criterion of a tail exponent below three~\cite{Tomberg:2025fku}. In the overdamped classical description the field never leaves the well, and the classical TCC excludes such a model outright. This does not rest on exact flatness: with a small tilt $\alpha$ toward the end surface, in units of the well width per e-fold, the classical duration in the well is $1/\alpha$, which the classical TCC excludes once it exceeds the remaining deadline, while the gap \eqref{eq:app-well-drift}, and with it the stochastic margin, changes by less than about $15\%$ wherever the stochastic TCC allows what the classical one excludes. Let a classical phase on a slope, with power spectrum $\PZ$ and duration $N_{\rm cl}$, end in the well, as in models of primordial black hole formation~\cite{Pattison:2017mbe,Ezquiaga:2019ftu,Vennin:2024yzl}. The reflecting top makes the two durations independent, and the first term of Eq.~\eqref{eq:wells-flat} with the generating function of the slope gives the worldline condition
\begin{equation}\label{eq:wells-wl}
\begin{gathered}
N_{\rm TCC}-N_{\rm cl}\;\geq\;\frac{4\mu^2}{\pi^2}\ln\frac{4}{\pi\delta}+\Delta N\,,\\[2pt]
\Delta N=\frac{N_{\rm cl}}{\Lambda_0\PZ}\Big(1-\sqrt{1-2\Lambda_0\PZ}\Big)-N_{\rm cl}\simeq\frac{\pi^2\PZ N_{\rm cl}}{8\mu^2}\,,
\end{gathered}
\end{equation}
valid for $\Lambda_0\leq1/(2\PZ)$. The margin, a few e-folds for $\mu^2$ of order one, is set by the gap; the noise of the classical phase adds only the small $\Delta N$, and a drift inside the well opens the gap and shortens the margin (Appendix~\ref{app:worldline-wells}).

The region reading is far more demanding. The classical phase turns one patch into $e^{3N_{\rm cl}}$ patches, each of which enters the well. If the well is eternal, each of them inflates forever with a probability bounded away from zero (Appendix~\ref{app:regions-pet}), and the region reading is violated with probability tending to one whatever the margin. Otherwise Eq.~\eqref{eq:stoch-union}, applied to the worldline bound behind Eq.~\eqref{eq:wells-wl}, gives the region condition
\begin{equation}\label{eq:wells-reg}
N_{\rm cl}+\Delta N\;\leq\;\Big(1-\frac{3}{\Lambda_0}\Big)N_{\rm TCC}-\frac{1}{\Lambda_0}\ln\frac{4}{\pi\delta}\,,
\end{equation}
the counterpart of Eq.~\eqref{eq:stoch-main} for a well: the deadline is shortened by the factor $1-3/\Lambda_0$, which vanishes at the eternal onset of the well. The condition is sufficient, and necessary up to a shift of order $1/\Lambda_0$ if the region reading in the well is proportional to $\mathbb E[Z]$, as on a slope. For a well somewhat smaller than the eternal size, a classical phase may then use only a fraction of the deadline, while per worldline it may use all but a few e-folds.

\section{Where the Stochastic TCC Matters}\label{sec:examples}
Applying the conditions to a model takes three steps. The end of inflation fixes the deadline. The model supplies the classical duration $N_{\rm cl}$ and the source of randomness: a power spectrum, the size $\mu^2$ of a well, or a decay rate. A reading and a confidence level then select the condition: Eq.~\eqref{eq:stoch-main} on a slope, or Eq.~\eqref{eq:stoch-general} per worldline where the power spectrum varies; Eq.~\eqref{eq:stoch-volume} for the volume; Eqs.~\eqref{eq:metastable-thresholds}--\eqref{eq:metastable-plane} with decay; Eqs.~\eqref{eq:wells-wl} and \eqref{eq:wells-reg} for a well. Table~\ref{tab:examples} applies them to five models at $\delta=10^{-3}$ and $H_{\rm end}=10^{-20}\Mp$, that is $N_{\rm TCC}\simeq46$.

\emph{The deadline.} The number of e-folds that the observable universe needs depends on the scale of inflation and on the history after it~\cite{Liddle:2003as}. The familiar fifty to sixty refer to inflation near $10^{16}\,$GeV, where the deadline is only $11.5$ e-folds. For rapid reheating, the two meet at the classical bound \eqref{eq:classical-bblv} on the scale of inflation, at about $45$ e-folds. Our deadline lies just above this point, with $V^{1/4}\simeq3\times10^8\,$GeV just below the bound: it is close to the shortest that inflation compatible with the TCC can have, and nearly all of it is needed. A prolonged reheating lowers the number needed by a third of the logarithm of the drop in temperature~\cite{Liddle:2003as}, and the shortest deadline by up to about $6$ e-folds. Lower scales lengthen the deadline and reduce the number of e-folds needed, up to $N_{\rm TCC}\simeq95$ for inflation at about $6\,$MeV, the lowest scale compatible with a reheating temperature of a few MeV, as nucleosynthesis requires~\cite{deSalas:2015glj}. For deadlines from about $39$ to $95$ e-folds no entry of Table~\ref{tab:examples} changes its verdict; toward longer deadlines the finite-deadline corrections shrink as $\ln(1/\delta)/N_{\rm TCC}$ and every entry moves toward its leading-order value. 

\begin{table}[htb]\centering\small
\begin{tabular}{lllll}\toprule
model & input & classical & stochastic, worldline & stochastic, region\\\midrule
(a) attractor & $\PZ=2.1\times10^{-9}$ & any margin & margin $\geq0.0012$ & margin $\geq0.0053$\\
(b) slope & $\PZ=0.1$ & $N_{\rm cl}<46.05$ & $N_{\rm cl}\leq38.1$ \ $(39.5^*)$ & $N_{\rm cl}\leq9.5$ \ $(10.6^*)$\\
(c) eternal slope & $\PZ=0.25$ & $N_{\rm cl}<46.05$\,$^\dagger$ & $N_{\rm cl}\leq33.4$ \ $(35.7^*)$ & $N_{\rm cl}\leq0.037^*$\\
(d) false vacuum & $\Gamma=0.2\,H^4$ & $\Gamma\geq0.005\,H^4$\,$^\dagger$ & $\Gamma\geq0.036\,H^4$ & $\Gamma\geq0.73\,H^4$\\
(e) well & $\mu^2=1/2$ & excluded & $N_{\rm cl}\leq44.6$ & $N_{\rm cl}\leq16.6$\\\bottomrule
\end{tabular}
\caption{The TCC for five models at $H_{\rm end}=10^{-20}\Mp$, that is $N_{\rm TCC}=46.05$, and $\delta=10^{-3}$: the longest classical duration, the shortest margin $N_{\rm TCC}-N_{\rm cl}$, or the smallest decay rate allowed. Entries follow from the sufficient conditions; those marked by an asterisk are exact, from the inverse Gaussian per worldline and from Eq.~\eqref{eq:app-fkpp} with binary splitting for a region. The decay rates are exact for patches that die independently, the region value for binary splitting. (a)~The CMB-normalized attractor; (b)~a slope below the onset of eternal inflation and (c)~one beyond it; (d)~a false vacuum below the eternal threshold $9H^4/(4\pi)=0.716\,H^4$; (e)~a slope with $\PZ\ll1$ that ends in a flat well with the gap $\Lambda_0=4.9$. Models (b)--(e) illustrate the readings and need not describe the phase that produced the observable universe. $^\dagger$Eternal inflation is ``marginally'' excluded by the slope bound with an undetermined coefficient in (c), and excluded in (d) if the logarithm of the lifetime bound is dropped, $\Gamma\gtrsim H^4$. The volume reading requires $\Gamma\geq0.75\,H^4$ in (d) and allows $N_{\rm cl}\leq21.1$ in (b).}
\label{tab:examples}
\end{table}

\emph{Unchanged.} On the attractor (a) the margins are thousandths of an e-fold in every reading. At this deadline, a model that respects the classical TCC by more than a hundredth of an e-fold respects the stochastic one in every reading, and the bound on the scale of inflation stands.

\emph{A model that fits the data.} The constant power spectrum of (a) has no tilt, but the conclusion holds as well for
models with the observed spectrum. A simple one is the hilltop
\begin{equation}\label{eq:examples-hilltop}
V=V_0\Big(1-\frac{\phi^2}{2f^2}\Big)\,,\qquad n_s=1-\frac{2}{f^2}\,,\qquad r=\frac{2V_0}{3\pi^2\PZ}\,,
\end{equation}
in which a fast waterfall ends inflation at a small field value, as in inverted hybrid inflation~\cite{Lyth:1996kt}. Its tilt does not depend on the number of e-folds before the end, which low scales make small, and its running is negligible; $f^2=67$ gives $n_s=0.97$, consistent with current measurements~\cite{Planck:2018jri,AtacamaCosmologyTelescope:2025blo}. At $N_{\rm TCC}=60$, within the allowed range, the observable universe needs $38$ e-folds, the CMB scales leave the horizon about $32$ e-folds before the end, and $r\sim10^{-44}$. Toward earlier times the power spectrum grows by the factor $e^{2/f^2}$ per e-fold, to at most $2.3$ times its CMB value at the start of a history that uses the whole deadline. Equation~\eqref{eq:stoch-general} then gives a margin of $0.0014$~e-folds per worldline and, with $\ln(1/\delta)$ replaced by $\ln(1/\delta)+3N_{\rm TCC}$ as in Eq.~\eqref{eq:stoch-union}, of $0.007$~e-folds for a region. The classical window between the $38$ e-folds needed and the deadline, $22$~e-folds wide, is untouched. The model assumes that inflation starts within this window, hundreds of e-folds of classical roll below the eternal top of the hill, and that the waterfall is fast; a slow one would act as a quantum well (Section~\ref{sec:wells}).

\emph{Stricter.} The slope (b) is close to the onset of eternal inflation without being eternal. The classical TCC and
the worldline reading allow most of the deadline, while the region reading lets the classical history use only between a fifth and a quarter of it. A classical history that uses half the deadline, which the classical TCC accepts with a wide margin, violates the region reading with a probability indistinguishable from one. The sufficient condition \eqref{eq:stoch-main} lies about one e-fold inside the exact one.

\emph{Decided by the reading.} On the eternal slope (c) the region reading allows only patches that start within a few hundredths of an e-fold of the end surface, in stochastic inflation effectively at it: this is the exclusion of eternal inflation with a definite threshold. Per worldline the same slope is allowed for every classical history that ends at least about $10$ e-folds before the deadline. The eternal false vacuum (d) splits in the same way: per worldline it is allowed by a wide margin, while the region and
volume readings require a decay rate above the eternal threshold. Here the classical TCC gives no definite verdict, since its exclusion of eternal inflation rests on an undetermined coefficient for (c) and on a dropped logarithm for (d)
(Section~\ref{sec:classical-consequences}).

\emph{More permissive.} The classical TCC excludes (e) outright, since the overdamped classical field never leaves the flat well. Per worldline the stochastic TCC asks only for a margin of one and a half e-folds, set by the gap of the well, and for a region it caps the classical phase at about a third of the deadline. An eternal well, $\mu^2=1$, needs a margin of three e-folds per worldline and is excluded for a region. Two other cases relax the classical TCC as well, with the caveats of Section~\ref{sec:stoch-sharp}: per worldline, a decay rate with $\kappa_{\rm d}N_{\rm TCC}>\ln(1/\delta)$ allows classical histories longer than the deadline, and the volume reading with the finite-history regulator accepts even a flat potential.

On the attractor, then, the classical TCC is an adequate proxy, with or without tilt. Near the onset of eternal inflation, for quantum wells and for false vacua it is not: it is too weak for a region near the onset, too strong for wells, and it leaves the fate of eternal inflation to an undetermined coefficient or a dropped logarithm. For such models a claim that the TCC is satisfied needs a reading, and the conditions derived here give it a definite content.

\section{Outlook: Rough Slopes and Several Fields}\label{sec:beyond}
Two extensions need stronger assumptions. We sketch them at the level of heuristics and leave their proof open.

\emph{Rough slopes.} Let the slope have a random part of short correlation length: in the units of Appendix~\ref{app:regions}, where a constant slope is $\beta=Q/2=(6\PZ)^{-1/2}$, let $\beta(X)=\bar\beta+\sigma_\beta B'(X)$ with $B$ a Brownian motion. Stretches on which the potential rises against the mean slope trap worldlines; a trap of depth $h$ in the potential $U=2\int\beta\,\dd X$ releases them at a rate of order $e^{-h}$, and the largest rise of $U$ ahead of a point is exponentially distributed with rate $\vartheta=\bar\beta/\sigma_\beta^2$. The spectral gap of a realization then has a power-law distribution near zero,
$\Pr[\Lambda_0<\lambda]\propto\lambda^{\vartheta}$, as for the random-force model~\cite{BOUCHAUD1990285,Comtet_1995}, and averaged over landscapes the violation probability of a worldline falls only as $N_{\rm TCC}^{-\vartheta}$, a power of the deadline rather than of $H_{\rm end}/\Mp$. This is the continuum form of the stable laws of hitting times of random walks in random environments~\cite{Kesten1975,10.2969/jmsj/04920189} and a Griffiths tail~\cite{PhysRevLett.23.17}; on a bounded field range a given landscape keeps an exponential tail with a random gap. For a region, traps deep enough to act as eternal wells (Section~\ref{sec:wells}) keep the violation probability above a floor that does not depend on the deadline. The effect needs relative slope fluctuations of order one near the onset of eternal inflation. For a slope normalized to the CMB a trap must reverse the slope within one correlation length, which has a Gaussian-small probability, and the worldline condition of the smooth slope applies.

\emph{Several fields.} If light fields select between classical continuations of a history~\cite{Guleryuz:2026ghb}, let ${\rm odds}^{\rm wl}$ be the ratio of the probabilities of two branches $\pm$ for one worldline, and ${\rm odds}^{\rm TCC}$ the same ratio among the histories that end before the deadline. The TCC reweights the branches by $\mathbb E[\mathbf 1_{\cN\leq N_{\rm TCC}}|+]/\mathbb E[\mathbf 1_{\cN\leq N_{\rm TCC}}|-]$, and when they differ by a small shift $\delta N$ of the duration, independent of the rest of it, to first order in the shift,
\begin{equation}\label{eq:beyond-transfer}
\frac{{\rm odds}^{\rm TCC}}{{\rm odds}^{\rm wl}}\simeq\exp\big[-\lambda_T\big(\langle\delta N\rangle_+-\langle\delta N\rangle_-\big)\big]\,,
\qquad\lambda_T=\frac{\dd\ln F(T)}{\dd T}\Big|_{T=N_{\rm TCC}}\,,
\end{equation}
with $F$ the distribution function of the branch-independent part of the duration. Volume weighting gives instead the factor $\exp[3(\langle\delta N\rangle_+-\langle\delta N\rangle_-)]$ for a deterministic shift~\cite{Linde:2010xz}. The TCC penalizes the longer branch, more strongly the earlier the deadline lies in the distribution of durations, and gives finite odds where volume weighting diverges, as when one branch crosses an eternal well.

\section{Conclusions}\label{sec:conclusions}
We have formulated the TCC in stochastic inflation. Its deadline becomes a fixed surface, the same for every history, and the conjecture a bound on the probability that a history reheats after it. What must reheat in time, one worldline, the region grown from one Hubble patch or the reheated volume, is a choice that the classical TCC never has to make. The region reading is the survival of branching Brownian motion with absorption up to the time $\ln(\Mp/H_{\rm end})$, which brings the survival theory of that process to bear on the TCC.

The central result is that each reading becomes one closed-form condition on the unused fraction of the deadline, Eqs.~\eqref{eq:stoch-main} and \eqref{eq:stoch-volume}: rigorous for a linear roll (within leading-order stochastic inflation and, for a region, the branching model), sharp at leading order in the deadline, and extended to a slowly varying power spectrum in the Bernstein form \eqref{eq:stoch-general}. Because the conditions are sharp, they decide what the classical TCC leaves open: the coefficient of the slope bound, whether eternal inflation is excluded and for which observer, and the bounds on the decay rate of a false vacuum. They also lift the classical TCC's exclusion of phases that end in quantum wells, per worldline and, below the eternal size, for a region. Table~\ref{tab:conclusions} collects the answers, and Section~\ref{sec:examples} applies them to models, among them one that fits the CMB data.

\emph{Stronger or weaker?} On a linear slope without vacuum decay, and at leading order on slowly varying slopes on which the duration is close to normal, the worldline and region readings are, short of the flat limit, never weaker than the classical TCC, and neither is the volume reading away from eternal inflation. They add a margin to the classical condition: a few standard deviations of the duration per worldline, and at least the fraction $\sqrt{6\PZ}$ of the deadline for a region. The stochastic TCC is weaker in three cases, each with the caveat stated in Section~\ref{sec:stoch-sharp}: a classical phase that ends in a flat or slightly tilted quantum well, which the overdamped classical TCC excludes, is allowed with a margin of a few e-folds per worldline; decay can end a worldline in time once $\kappa_{\rm d}N_{\rm TCC}>\ln(1/\delta)$; and the volume reading with the finite-history regulator can accept eternal models.

\emph{What does it allow or forbid?} On the CMB-normalized attractor the margins are a fraction of an e-fold, and the bounds on the scale of inflation and on $r$ hold in every reading. Where the noise competes with the drift the reading decides. The region reading forbids eternal inflation on a linear slope for every classical history that uses a finite fraction $c$ of the deadline, and its slope bound $|V'|>CV^{3/2}$ has, at leading order, the coefficient $C=1/[\sqrt2\,\pi(1-c)]$: the known onset value as $c\to0$, and for finite $c$ the exclusion of a band of slopes that are not eternal. For a false vacuum the region and volume readings require a decay rate somewhat above the eternal threshold, the region reading, for binary splitting, leaving only a window of relative width $\delta$ below it at deadlines beyond $(1/\delta-1)/3$, which for $\delta=10^{-3}$ is longer than the deadline of the present vacuum. Per worldline the TCC forbids neither form of eternal inflation: the Hubble patch of one observer reheats or decays in time with high probability, while the inflating volume grows forever. The ``marginal'' exclusion of eternal inflation by the classical TCC thus resolves into a choice between a single-observer and a global formulation, and, for a false vacuum, into whether the logarithm of the
lifetime bound is kept. 

\begin{table}[t]\centering\small
\setlength{\tabcolsep}{4pt}
\begin{tabular}{>{\raggedright\arraybackslash}p{0.18\textwidth}>{\raggedright\arraybackslash}p{0.21\textwidth}>{\raggedright\arraybackslash}p{0.24\textwidth}>{\raggedright\arraybackslash}p{0.28\textwidth}}\toprule
 & classical & stochastic, worldline & stochastic, region\\\midrule
slow roll, $N_{\rm TCC}\to\infty$ & $c<1$ & $c<1$ & $c\leq1-\sqrt{6\PZ}$\\[2pt]
scale of inflation, $r$ & Eq.~\eqref{eq:classical-bblv} & unchanged & unchanged\\[2pt]
eternal slow roll & ``marginally'' excluded, $C$ undetermined & allowed & excluded where $P_{\rm et}>\delta$, hence for every $c>0$ at leading order;
slope bound $C=1/[\sqrt2\,\pi(1-c)]$\\[2pt]
false vacuum, $\Gamma/H^4\geq$ & $3/(4\pi N_{\rm TCC})$; eternal allowed & $3\ln(1/\delta)/(4\pi N_{\rm TCC})$;
eternal allowed & about $9/(4\pi)$; eternal excluded if $P_{\rm et}>\delta$\\[2pt]
classical phase ending in a flat well & excluded (duration diverges) & margin $\dfrac{4\mu^2}{\pi^2}\ln\dfrac{4}{\pi\delta}$ &
$c\lesssim1-3/\Lambda_0$; eternal wells excluded\\\bottomrule
\end{tabular}
\caption{Classical and stochastic consequences of the TCC at confidence $1-\delta$, with $c=N_{\rm cl}/N_{\rm TCC}$ the
fraction of the deadline used by the classical history, on a slope without decay unless a decay rate is shown. The
volume reading gives $c<\sqrt{1-6\PZ}$ on a slope and $\Gamma/H^4\geq9/(4\pi)+3\ln(1/\delta)/(4\pi N_{\rm TCC})$ for a false vacuum, and depends on the measure where inflation is eternal. In the last row $\Lambda_0=\pi^2/(4\mu^2)$ is the gap of the well (Section~\ref{sec:wells}).}
\label{tab:conclusions}
\end{table}

Four questions are left open. The most important is whether the spectral gap, and with it every exponent above, is stable
under quantum corrections to the stochastic description~\cite{Gorbenko:2019rza,Mirbabayi:2019qtx,Baumgart:2019clc,Cohen:2020php,Green:2026nnw}. The others concern realistic wells
for primordial black holes, entered through a phase of ultra-slow roll; power spectra that vary on the scale of the margin, are rough, or grow up the potential into an eternal regime; and several fields with correlated branches. Each has a first-passage formulation in which the deadline enters in the same way.

\section*{Acknowledgments}
The contribution of O.G. was supported in part by the Istanbul Technical University Research Fund under grant number 2026-49066. Claude (Anthropic) was used to assist with code review and debugging, consistency checks, and manuscript editing. The author takes full responsibility for the content of the article.

\appendix
\section{Derivations for a Worldline}\label{app:worldline}
\subsection{Spectral form and the inverse-Gaussian law}\label{app:worldline-spectral}
With the speed measure $m(\phi)=e^{\int^\phi b/D}/D$ the generator takes the Sturm--Liouville form $\mathcal Lf=m^{-1}(Dm\,f')'$. It is symmetric in $L^2(m\,\dd\phi)$ on functions that vanish on the end surface and have vanishing derivative at a reflecting wall, and on a bounded interval its eigenvalues, $\mathcal L\psi_n=-\Lambda_n\psi_n$, are discrete and simple, with $0<\Lambda_0<\Lambda_1<\cdots$. The survival function is
\begin{equation}\label{eq:app-spectral}
\mathcal S(N|\phi)=\sum_na_n\psi_n(\phi)\,e^{-\Lambda_nN}\,,\qquad a_n=\frac{\int m\psi_n\,\dd\phi}{\int m\psi_n^2\,\dd\phi}\,,
\end{equation}
which at the deadline is Eq.~\eqref{eq:stoch-spectral}. In slow roll the speed measure, $m\propto e^{S}/V$ with $S=24\pi^2/V$ the de Sitter entropy, is the weight of the stationary distribution~\cite{Starobinsky:1986fx,Starobinsky:1994bd}.

For a linear roll let $y=\phi-\phi_{\rm end}$ start at $d>0$ and move with constant drift $-|b|$ and diffusion coefficient $D$. The method of images gives
\begin{equation}\label{eq:app-ig}
\mathcal S(N)=\Phi\Big(\frac{N_{\rm cl}-N}{\sqrt{\PZ N}}\Big)-e^{2N_{\rm cl}/\PZ}\,\Phi\Big(-\frac{N_{\rm cl}+N}{\sqrt{\PZ N}}\Big)\,,
\qquad N_{\rm cl}=\frac{d}{|b|}\,,\quad\PZ=\frac{2D}{b^2}\,,
\end{equation}
the inverse Gaussian with mean $N_{\rm cl}$ and variance $\PZ N_{\rm cl}$. Its moment generating function is the solution of $Du''-|b|u'+su=0$ that equals one on the end surface and grows most slowly with $y$,
\begin{equation}\label{eq:app-mgf}
\mathbb E\big[e^{s\cN}\big]=\exp\Big[\frac{N_{\rm cl}}{\PZ}\Big(1-\sqrt{1-2\PZ s}\Big)\Big]\,,\qquad s\leq\frac{1}{2\PZ}=\Lambda_0\,.
\end{equation}
It is finite at $s=3$ if and only if $\PZ\leq1/6$, the onset of eternal inflation on a linear slope~\cite{Creminelli:2008es}.

\subsection{The Chernoff and Bernstein bounds}\label{app:worldline-chernoff}
For $0\leq s\leq\Lambda_0$ Markov's inequality gives $\Pr[\cN>T]\leq e^{-sT}\mathbb E[e^{s\cN}]$. With Eq.~\eqref{eq:app-mgf} the exponent is smallest where $\sqrt{1-2\PZ s}=N_{\rm cl}/T$, inside the allowed range for $T\geq N_{\rm cl}$, and there it equals
\begin{equation}\label{eq:app-chernoff}
-\frac{T}{2\PZ}\Big(1-\frac{N_{\rm cl}^2}{T^2}\Big)+\frac{N_{\rm cl}}{\PZ}\Big(1-\frac{N_{\rm cl}}{T}\Big)=-\frac{(T-N_{\rm cl})^2}{2\PZ T}\,,
\end{equation}
which is Eq.~\eqref{eq:stoch-chernoff}. This is the exact large-deviation exponent of the inverse Gaussian: at fixed $N_{\rm cl}/T$ between zero and one the tail carries only an extra prefactor of order $T^{-1/2}$. The volume-weighted duration is inverse Gaussian as well, with mean $N_{\rm cl}/\sqrt{1-\chi}$ and power spectrum $\PZ/(1-\chi)$ (Appendix~\ref{app:regions-vol}), which gives Eq.~\eqref{eq:stoch-volume}.

\emph{Varying power spectrum.} Let $y$ be the distance to the end surface along the classical path. The generating function $u_s=\mathbb E[e^{s\cN}]$ still solves $Du''-|b|u'+su=0$ with $u_s=1$ on the end surface. At leading order in the slow variation of $b$ and $D$ the second derivative of $\Theta=\ln u_s$ is negligible against the square of the first. Then $D\Theta'^2-|b|\Theta'+s=0$, and integrating along the path, with $\dd N=\dd y/|b|$, gives
\begin{equation}\label{eq:app-adiabatic}
\ln\mathbb E\big[e^{s\cN}\big]\simeq K(s)=\int_0^{N_{\rm cl}}\frac{1-\sqrt{1-2s\PZ}}{\PZ}\,\dd N
=sN_{\rm cl}+\int_0^{N_{\rm cl}}\frac{2s^2\,\PZ}{\big(1+\sqrt{1-2s\PZ}\big)^2}\,\dd N\,.
\end{equation}
The weight that multiplies $\PZ$ in the last integral grows with $\PZ$. Bounding it by its value at the largest power spectrum $\mathcal P_{\max}$ gives, with $\sigma_N^2=\int\PZ\,\dd N$,
\begin{equation}\label{eq:app-K}
K(s)\leq s\,(N_{\rm cl}-M)+\frac{M}{\mathcal P_{\max}}\Big(1-\sqrt{1-2s\mathcal P_{\max}}\Big)\,,\qquad M=\frac{\sigma_N^2}{\mathcal P_{\max}}\leq N_{\rm cl}\,,
\end{equation}
the generating function of a constant $N_{\rm cl}-M$ plus a linear roll of duration $M$ and power spectrum $\mathcal P_{\max}$. Equation~\eqref{eq:app-chernoff} for that roll, evaluated at $T-N_{\rm cl}+M$, gives Eq.~\eqref{eq:stoch-bernstein}.

The adiabatic form neglects a correction to the mean duration of order $\mathcal P_{\max}$ times the logarithmic change of the
slope, $-\frac12\int\PZ\,\dd\ln|b|$ integrated outward from the end surface. It therefore holds for margins large compared with $\mathcal P_{\max}$, the moderate deviations that the worldline reading probes. In the far tail the generating function is limited instead by the spectral gap of the whole field range that fluctuations can reach, which may lie below $1/(2\mathcal P_{\max})$ if the power spectrum grows up the potential. We compared the bound with $10^5$ Langevin paths in $V\propto\phi^2$, $V\propto\phi^4$ and a plateau $V\propto(1-e^{-\sqrt{2/3}\phi})^2$, with $\PZ$ up to $0.14$ at the initial field value and margins from half an e-fold to two. The bound exceeded the simulated violation probability by factors between about three and fifty, while the Gaussian estimate without the term $\mathcal P_{\max}t$ fell below it at the larger margins of the shorter histories, in some cases by more than two orders of magnitude.

\subsection{Flat wells}\label{app:worldline-wells}
In units of the well width, $x=(\phi-\phi_{\rm end})/\Delta\phi$, the generator is $\mu^{-2}\partial_x^2$ on $0<x<1$, absorbing at $x=0$ and reflecting at $x=1$. Its eigenfunctions and eigenvalues are
\begin{equation}\label{eq:app-well-modes}
\psi_n=\sin k_nx\,,\qquad\Lambda_n=\frac{k_n^2}{\mu^2}\,,\qquad k_n=\frac{(2n+1)\pi}{2}\,,
\end{equation}
and the initial datum $\mathcal S=1$ has the projections $4/[(2n+1)\pi]$, which gives Eq.~\eqref{eq:wells-flat} at $x=1$. The series alternates with decreasing terms, so it is bounded by its first term, $(4/\pi)e^{-\Lambda_0N}$, which gives Eq.~\eqref{eq:wells-wl}. Where the first term equals $\delta$ the next is smaller by the factor $\frac13(\pi\delta/4)^8$, so the bound also gives the exact margin.

A constant drift $\alpha$ toward the end surface, in units of the well width per e-fold, turns the eigenfunctions into $e^{\gamma x}\sin kx$, with $\gamma=\alpha\mu^2/2$ and $k$ fixed by the reflecting wall, $\gamma\sin k+k\cos k=0$. It
opens the gap,
\begin{equation}\label{eq:app-well-drift}
\Lambda_0=\frac{k^2+\gamma^2}{\mu^2}\simeq\frac{\pi^2}{4\mu^2}+\alpha\,,
\end{equation}
the approximation holding for a small drift.

\section{Derivations for Regions and Volumes}\label{app:regions}
For a linear roll, we measure the distance from the end surface in units in which the diffusion coefficient equals the volume growth rate,
\begin{equation}\label{eq:app-units}
\begin{gathered}
X=\sqrt{\frac3D}\,(\phi-\phi_{\rm end})\,,\qquad\dd X=-3Q\,\dd N+\sqrt6\,\dd W_N\,,\qquad
Q=\frac{|b|}{\sqrt{3D}}=\Big(\frac{2}{3\PZ}\Big)^{1/2}\,,\\[2pt]
\mathcal L=3\big(\partial_X^2-Q\,\partial_X\big)\,,\qquad N_{\rm cl}=\frac{X}{3Q}\,,\qquad\Lambda_0=\frac{3Q^2}{4}\,,
\end{gathered}
\end{equation}
so that the onset of eternal inflation without decay is $Q=2$. Patches move as independent worldlines, split in two at the rate $3$, die at the rate $\kappa_{\rm d}$, and are removed at $X=0$.

\subsection{Many-to-one lemma and the equation for the region}\label{app:regions-fkpp}
The mean number of patches obeys $\partial_T\mathbb E_X[Z_T]=(\mathcal L+3-\kappa_{\rm d})\,\mathbb E_X[Z_T]$, so that $\mathbb E[Z_T]=e^{(3-\kappa_{\rm d})T}\mathcal S(T|X)=e^{3T}P^{\rm wl}$; only the mean growth enters, whatever the branching law. In the first interval $\dd T$ a patch splits with probability $3\,\dd T$ or dies with probability $\kappa_{\rm d}\dd T$, and the subtrees that follow are independent. Conditioning on that interval shows that $u=P^{\rm reg}$ solves the Fisher--KPP equation~\cite{https://doi.org/10.1111/j.1469-1809.1937.tb02153.x,KPP1937}
\begin{equation}\label{eq:app-fkpp}
\partial_Tu=\mathcal Lu+3u(1-u)-\kappa_{\rm d}u\,,\qquad u(0,X)=1\,,\qquad u(T,0)=0\,.
\end{equation}
The events $\{Z_T>0\}$ decrease in $T$ and their intersection is survival forever, so $P^{\rm reg}$ decreases to the stationary solution $P_{\rm et}$.

Near the end surface, $P_{\rm et}$ is small and solves the linearized stationary equation. Its solution that vanishes at $X=0$ is
\begin{equation}\label{eq:app-pet-lin}
P_{\rm et}\propto e^{QX/2}\sin\omega X\,,\qquad\omega^2=1-\frac{\Lambda_0+\kappa_{\rm d}}{3}\,,
\end{equation}
oscillatory only on the eternal side. It grows until it saturates within the correlation length $\pi/\omega$, which diverges at the onset: $P_{\rm et}$ is of order one beyond it and suppressed roughly as $e^{-Q(\pi/\omega-X)/2}\sin\omega X$ closer to the end surface.

\subsection{The volume reading as a slower roll}\label{app:regions-vol}
The duration of a patch at $X$ has the density
\begin{equation}\label{eq:app-density}
f_Q(N)=\frac{X}{\sqrt{12\pi N^3}}\,\exp\Big[-\frac{(X-3QN)^2}{12N}\Big]\,.
\end{equation}
Expanding the square shows that the volume weight $e^{\rho N}$, with $\rho=3-\kappa_{\rm d}$, maps it into the same family,
\begin{equation}\label{eq:app-tilt}
e^{\rho N}f_Q(N)=e^{X(Q-\tilde Q)/2}f_{\tilde Q}(N)\,,\qquad\frac{3\tilde Q^2}{4}=\frac{3Q^2}{4}-\rho\,,\qquad
\tilde Q=Q\sqrt{1-\chi}\,.
\end{equation}
For $\chi<1$ the volume-weighted duration is therefore inverse Gaussian with mean $N_{\rm cl}/\sqrt{1-\chi}$ and power spectrum $2/(3\tilde Q^2)=\PZ/(1-\chi)$, and the normalization is $\mathbb E[e^{\rho\cN}]=e^{X(Q-\tilde Q)/2}$. For $\chi>1$ the tilted density grows as $N^{-3/2}e^{(3-\Lambda_0-\kappa_{\rm d})N}$, and the normalization diverges.

\subsection{Eternal inflation as positive survival}\label{app:regions-pet}
On the half-line, if $\Lambda_0+\kappa_{\rm d}\geq3$, then $\mathbb E[Z_T]=e^{(3-\kappa_{\rm d})T}\mathcal S(T)\to0$ and $P_{\rm et}=0$. If $\Lambda_0+\kappa_{\rm d}<3$, remove also the patches at $X=\ell$, with $\ell>X$ so large that $\lambda_\ell\equiv3-\Lambda_0-\kappa_{\rm d}-3\pi^2/\ell^2>0$. The positive function $\psi_\ell(X)=e^{QX/2}\sin(\pi X/\ell)$ obeys $\mathcal L\psi_\ell=-(\Lambda_0+3\pi^2/\ell^2)\psi_\ell$. The sum $\Sigma_T$ of $\psi_\ell$ over the patches in the strip then has, by the first- and second-moment formulas of binary branching~\cite{10.1214/15-AIHP714} and
$\psi_\ell^2\leq\|\psi_\ell\|_\infty\psi_\ell$,
\begin{equation}\label{eq:app-moments}
\mathbb E\,\Sigma_T=e^{\lambda_\ell T}\psi_\ell(X)\,,\qquad \mathbb E\,\Sigma_T^2\leq\|\psi_\ell\|_\infty\psi_\ell(X)\Big[e^{\lambda_\ell T}+\frac{6}{\lambda_\ell}\,e^{2\lambda_\ell T}\Big]\,.
\end{equation}
The Paley--Zygmund inequality, $\Pr[\Sigma_T>0]\geq(\mathbb E\,\Sigma_T)^2/\mathbb E\,\Sigma_T^2$, gives
\begin{equation}\label{eq:app-pz}
\Pr[Z_T>0]\geq\frac{\psi_\ell(X)}{(1+6/\lambda_\ell)\,\|\psi_\ell\|_\infty}>0\qquad\text{for all }T\,,
\end{equation}
so $P_{\rm et}(X)>0$. Only the positivity of the Dirichlet ground state was used. On a bounded domain and for any slope, therefore, $P_{\rm et}>0$ whenever the principal eigenvalue of $-\mathcal L+\kappa_{\rm d}$ lies below three, and $P_{\rm et}=0$ when it lies strictly above, since then $\mathbb E[Z_T]\to0$; the critical case on a bounded domain is not needed here.

\subsection{The region condition is sharp}\label{app:regions-front}
The argument combines a first-moment bound with a comparison with the front of branching Brownian motion. Let $\kappa_{\rm d}<3$, $T=N_{\rm TCC}$ and $X=3QcT$. Without a wall the front of the population moves at the speed $3Q_{\rm f}$ relative to the mean drift~\cite{Mckean1975ApplicationOB,VANSAARLOOS_2003}, with
\begin{equation}\label{eq:app-front}
Q_{\rm f}=2\Big(1-\frac{\kappa_{\rm d}}{3}\Big)^{1/2}=Q\sqrt\chi\,.
\end{equation}

\emph{Below the threshold.} If $c<1-\sqrt\chi$, Eqs.~\eqref{eq:stoch-union} and \eqref{eq:stoch-chernoff}, with the decay factor $e^{-\kappa_{\rm d}T}$, give $P^{\rm reg}\leq\mathbb E[Z_T]\to0$ exponentially. At $c=1-\sqrt\chi$ the exponent vanishes, and the inverse-Gaussian prefactor gives $\mathbb E[Z_T]=O(T^{-1/2})\to0$.

\emph{Above the threshold.} Let $1-\sqrt\chi<c\leq1$; larger $c$ follow, since a coupling shows that $P^{\rm reg}$ does not decrease with the initial distance. Choose $\varepsilon$ with $0<\varepsilon<\min\{Qc,\,Q(c-1+\sqrt\chi)\}$, so that $Q'\equiv Q(1-c)+\varepsilon$ obeys $0<Q'<Q_{\rm f}$. In the coordinate $Y=X+3QN$ the patches perform driftless Brownian motions, the absorbing wall is the line $Y=3QN$, and the initial patch starts at $Y=3QcT$. The line
\begin{equation}\label{eq:app-line}
Y_*(N)=3(Qc-\varepsilon)T+3Q'N\,,\qquad Y_*(N)-3QN=3(Qc-\varepsilon)(T-N)\geq0\,,
\end{equation}
lies above the wall until the deadline, so a patch that stays above $Y_*$ has not been absorbed. Hence $P^{\rm reg}(T)\geq\tilde P(3\varepsilon T)$, with $\tilde P(x)$ the probability that a population with drift $-3Q'$, started at a distance $x$ above an absorbing wall, survives forever. Since $3Q'^2/4+\kappa_{\rm d}<3Q_{\rm f}^2/4+\kappa_{\rm d}=3$, Appendix~\ref{app:regions-pet} gives $\tilde P(x)>0$.

\emph{The limit.} The function $\tilde P$ is nondecreasing and tends to $1-\kappa_{\rm d}/3$ as $x\to\infty$. On the event that the population without a wall survives, which has probability $1-\kappa_{\rm d}/3$, its size at a time $t_0$ is large when $t_0$ is, and all its patches lie within some distance $Kt_0$ of the start with probability close to one; each of them founds an independent population that survives the wall with probability at least $\tilde P(x-Kt_0)$. Since $P^{\rm reg}$ cannot exceed the survival probability without a wall, $P^{\rm reg}\to1-\kappa_{\rm d}/3$, which is Eq.~\eqref{eq:stoch-step}. The argument holds for any branching law with finite variance, with $1-\kappa_{\rm d}/3$ replaced by the survival probability of its Galton--Watson process,  and for $\kappa_{\rm d}\geq3$ the limit vanishes for every $c$.

\subsection{Finite histories}\label{app:regions-fin}
Here the volume of a patch grows continuously, $v\to ve^{3\dd N}$, rather than by splitting. A region of volume $v$ at $X$ then has the transform $g(X;z)^v$, where $g=\mathbb E_X[e^{-z\mathcal V}]$ and $\mathcal V$ is the reheated volume of one patch, and conditioning on the first step gives~\cite{Winitzki:2008yb,Winitzki:2008ph}
\begin{equation}\label{eq:app-g}
g''-Qg'+g\ln g=0\,,\qquad g(0;z)=e^{-z}\,.
\end{equation}
As $z\to0^+$, $q=-\ln g$ tends to $q_\infty=-\ln(1-P_{\rm et})$, with $P_{\rm et}$ the probability of eternal inflation in this model.

\emph{The line of descent.} Differentiating Eq.~\eqref{eq:app-g} in $z$ shows that
$h_z=-\partial_zg=\mathbb E[\mathcal Ve^{-z\mathcal V}]$ solves $(\mathcal L+3-3q)h_z=0$. By the Feynman--Kac formula the line of descent of a point of the reheated volume is thus a worldline weighted by $e^{3\cN}$ and killed at the rate $3q$, each subtree that branches off contributing its transform; this is the spine decomposition of branching diffusions~\cite{hardy2006newformulationspineapproach,10.1214/15-AIHP714}. Differentiating Eq.~\eqref{eq:app-g} in $X$ shows that $g'$ solves the same equation; as $z\to0^+$ both solutions decay at large $X$, so $h_0=P_{\rm et}'/P_{\rm et}'(0)$. The violating fraction
among volume-weighted finite histories is then
\begin{equation}\label{eq:app-fin}
P^{\rm fin}(T)=\frac{1}{h_0(X)}\,\mathbb E_X\Big[\mathbf 1_{\cN>T}\,e^{\int_0^T(3-3q_\infty)\dd N}\,h_0(X_T)\Big]
\propto e^{-\Lambda_{\rm fin}T}\,,
\end{equation}
with $\Lambda_{\rm fin}$ the Dirichlet ground-state energy of $-(\mathcal L+3-3q_\infty)$ and $\psi_{\rm fin}$ its ground state. Comparing the eigenvalue equation with that of $h_0$, through $\mathcal L=3e^{QX}\partial_X(e^{-QX}\partial_X)$, leaves only the boundary term,
\begin{equation}\label{eq:app-green}
\Lambda_{\rm fin}=\frac{3\,h_0(0)\,\psi_{\rm fin}'(0)}{\int_0^\infty e^{-QX}\psi_{\rm fin}h_0\,\dd X}>0\,.
\end{equation}

\emph{Flat potential.} For $Q=0$ the solution of Eq.~\eqref{eq:app-g} at $z\to0^+$, selected by its growth at large $X$ and confirmed by shooting, is
\begin{equation}\label{eq:app-qflat}
q_\infty=\frac{X^2}{4}+\frac{X}{\sqrt2}\,,
\end{equation}
which solves $q'=(q+\tfrac12)^{1/2}$. With $\tilde X=X+\sqrt2$ the operator becomes $3(-\partial_{\tilde X}^2+\tilde X^2/4-3/2)$, whose decaying eigenfunctions are the parabolic cylinder functions $D_\nu(\tilde X)$, with eigenvalues $3(\nu-1)$. The smallest root of the Dirichlet condition $D_\nu(\sqrt2)=0$ gives $\Lambda_{\rm fin}=4.6116$.

\emph{Near the onset.} Within a correlation length $q_\infty$ is small, and $h_0\propto e^{QX/2}\sin(\omega X+\arcsin\omega)$ with $\omega^2=1-Q^2/4$. Its zero $X_h=(\pi-\arcsin\omega)/\omega$ acts as a wall where the nonlinear region begins, and the ground state of the strip $(0,X_h)$ has the energy
\begin{equation}\label{eq:app-onset}
\Lambda_{\rm fin}\simeq3\Big(\frac{\pi^2}{X_h^2}-\omega^2\Big)\simeq\frac6\pi\,\omega^3\,,
\end{equation}
the leading-order onset law of Section~\ref{sec:slope-regulators}.

\section{The Simulated Universes}\label{app:lattice}
The diagrams of Fig.~\ref{fig:penrose}(b)--(d) come from one realization of stochastic inflation on a three-dimensional comoving lattice in the separate-universe approximation, in the spirit of earlier lattice simulations~\cite{Linde:1993xx,Vanchurin:1999iv,Mizuguchi:2024kbl}.

\emph{Lattice.} Lengths are in units of the classical end-of-inflation Hubble radius, so that the comoving Hubble wavenumber is $k_H(N)=e^{N-N_{\rm cl}}$ and the initial Hubble patch has radius $e^{N_{\rm cl}}$. A periodic box of side $20$ with $240^3$ points holds the initial patch, with $N_{\rm cl}=1.5$, and its surroundings. In the units of Eq.~\eqref{eq:app-units} the field is $X=3Q(N_{\rm cl}-N)+\Xi(\mathbf x,N)$, where the Gaussian field $\Xi$ receives, in each step of $0.03$ e-folds, the Fourier modes with $k_H(N)\leq|\mathbf k|<k_H(N+0.03)$ of a scale-invariant spectrum with variance $6$ per e-fold; modes longer than the box enter as a common kick. Every point thus performs the Langevin motion of Eq.~\eqref{eq:stoch-langevin}, and neighboring points are correlated on the Hubble scale at which their modes froze. A point reheats at its first passage through $X=0$: at the end of a step or, with the crossing probability of the Brownian bridge between the two ends, inside it. The uniform variate of this test is shared on blocks of the comoving Hubble size of the step, since the noise inside a step comes from modes of that size, and the crossing time is interpolated inside the step. The same noise drives the three slopes $\PZ=0.01$, $0.08$ and $0.25$.

\emph{Below the lattice scale.} The lattice resolves the comoving Hubble scale until $k_H$ reaches the Nyquist wavenumber, $3.6$ e-folds after the classical end. Points that still inflate then continue with the exact statistics of each point, while the correlations between points below the lattice scale are approximate. Along the drawn ray the noise is shared on blocks of length $1/k_H$ until the points reheat, so that the ray follows single worldlines; in panel (d) the ray therefore reheats even at the center, whose subtree never stops inflating. The latest reheating inside a lattice cell is the extinction time of its subtree of Hubble patches, computed by iterating the generating function of an eightfold split every $\ln2$ e-folds. Shared histories make the reheating surface along the ray nearly constant on the Hubble scale at reheating, hence its steps.

\emph{Diagrams.} Each diagram is centered on the point of the initial patch, $|\mathbf x|<e^{N_{\rm cl}}$, that reheats last, in panel (d) on a cell whose subtree never stops inflating. The ray leaves it in a fixed direction and runs to $r=30$ through the periodic box, keeping a distance larger than $11$ from the periodic images of its origin. Beyond $r=30$, the last few percent of each diagram toward $i^0$, the classical surface is drawn. The reheating times along the ray are mapped to $\eta=-e^{-(N-N_{\rm cl})}$ in the flat slicing of the background, compactified with the null coordinates $\arctan[(\eta-r)/L]$ and $\arctan[(\eta+r)/L]$, $L=1.5$, in units of the classical end-of-inflation Hubble radius.

\emph{Other numerics.} The values of $P^{\rm reg}$ quoted in the text are obtained by solving Eq.~\eqref{eq:app-fkpp} by the method of lines, with grid spacing $0.025$ to $0.05$, a backward-differentiation integrator and a far wall whose position and boundary value do not affect the results. Refining the grid changes the results by less than $10^{-3}$ relative for values of order one and by a few percent for the value $10^{-7}$ of Fig.~\ref{fig:penrose}(b), and a direct simulation of the branching process agrees within its errors. The binary-splitting bound follows from the root of Eq.~\eqref{eq:metastable-region}; the bounds for the Poisson law are obtained by iterating its generating function once per e-fold, over $11$ and $12$ e-folds for $N_{\rm TCC}=11.5$. At the onset of eternal inflation, $\Lambda_0+\kappa_{\rm d}=3$, Eq.~\eqref{eq:app-fkpp} is solved for $e^{-QX/2}u$, which removes the linear terms other than diffusion and so does not shift the onset; integrated to $N_{\rm TCC}=3000$, the solution reproduces the exponent of Eq.~\eqref{eq:slope-kesten} and the profile $Xe^{QX/2}$ to better than one percent for $\kappa_{\rm d}=0$ and $1$.

\bibliography{refs}

@article{BOUCHAUD1990285,
title = {Classical diffusion of a particle in a one-dimensional random force field},
journal = {Annals of Physics},
volume = {201},
number = {2},
pages = {285-341},
year = {1990},
issn = {0003-4916},
doi = {https://doi.org/10.1016/0003-4916(90)90043-N},
url = {https://www.sciencedirect.com/science/article/pii/000349169090043N},
author = {J.P Bouchaud and A Comtet and A Georges and P {Le Doussal}}
}

@article{Comtet_1995,
   title={Localization Properties in One-Dimensional Disordered Supersymmetric Quantum Mechanics},
   volume={239},
   ISSN={0003-4916},
   url={http://dx.doi.org/10.1006/aphy.1995.1037},
   DOI={10.1006/aphy.1995.1037},
   number={2},
   journal={Annals of Physics},
   publisher={Elsevier BV},
   author={Comtet, A. and Desbois, J. and Monthus, C.},
   year={1995},
   month=May, pages={312–350} }

@article{Kesten1975,
author = {Kesten, Harry and Kozlov, M. V. and Spitzer, Frank},
journal = {Compositio Mathematica},
number = {2},
pages = {145-168},
publisher = {Noordhoff International Publishing},
title = {A limit law for random walk in a random environment},
url = {http://eudml.org/doc/89251},
volume = {30},
year = {1975}
}

@article{10.2969/jmsj/04920189,
author = {Kiyoshi KAWAZU and Hiroshi TANAKA},
title = {{A diffusion process in a Brownian environment with drift}},
volume = {49},
journal = {Journal of the Mathematical Society of Japan},
number = {2},
publisher = {Mathematical Society of Japan},
pages = {189 -- 211},
year = {1997},
doi = {10.2969/jmsj/04920189},
URL = {https://doi.org/10.2969/jmsj/04920189}
}

@article{PhysRevLett.23.17,
  title = {Nonanalytic Behavior Above the Critical Point in a Random Ising Ferromagnet},
  author = {Griffiths, Robert B.},
  journal = {Phys. Rev. Lett.},
  volume = {23},
  issue = {1},
  pages = {17--19},
  numpages = {0},
  year = {1969},
  month = {Jul},
  publisher = {American Physical Society},
  doi = {10.1103/PhysRevLett.23.17},
  url = {https://link.aps.org/doi/10.1103/PhysRevLett.23.17}
}

@article{Guleryuz:2026ghb,
    author = "Guleryuz, Omer",
    title = "{Quantum Selection of Classical Histories}",
    eprint = "2608.27388",
    archivePrefix = "arXiv",
    primaryClass = "hep-th",
    month = "8",
    year = "2026"
}

@article{Linde:2010xz,
    author = "Linde, Andrei and Noorbala, Mahdiyar",
    title = "{Measure Problem for Eternal and Non-Eternal Inflation}",
    eprint = "1006.2170",
    archivePrefix = "arXiv",
    primaryClass = "hep-th",
    doi = "10.1088/1475-7516/2010/09/008",
    journal = "JCAP",
    volume = "09",
    pages = "008",
    year = "2010"
}

@article{Bousso:2002ju,
    author = "Bousso, Raphael",
    title = "{The Holographic principle}",
    eprint = "hep-th/0203101",
    archivePrefix = "arXiv",
    reportNumber = "NSF-ITP-02-17, NSF-ITP-02-17",
    doi = "10.1103/RevModPhys.74.825",
    journal = "Rev. Mod. Phys.",
    volume = "74",
    pages = "825--874",
    year = "2002"
}

@article{Bedroya:2019snp,
    author = "Bedroya, Alek and Vafa, Cumrun",
    title = "{Trans-Planckian Censorship and the Swampland}",
    eprint = "1909.11063",
    archivePrefix = "arXiv",
    primaryClass = "hep-th",
    doi = "10.1007/JHEP09(2020)123",
    journal = "JHEP",
    volume = "09",
    pages = "123",
    year = "2020"
}

@article{Arkani-Hamed:2007ryv,
    author = "Arkani-Hamed, Nima and Dubovsky, Sergei and Nicolis, Alberto and Trincherini, Enrico and Villadoro, Giovanni",
    title = "{A Measure of de Sitter entropy and eternal inflation}",
    eprint = "0704.1814",
    archivePrefix = "arXiv",
    primaryClass = "hep-th",
    doi = "10.1088/1126-6708/2007/05/055",
    journal = "JHEP",
    volume = "05",
    pages = "055",
    year = "2007"
}

@article{Dubovsky:2008rf,
    author = "Dubovsky, Sergei and Senatore, Leonardo and Villadoro, Giovanni",
    title = "{The Volume of the Universe after Inflation and de Sitter Entropy}",
    eprint = "0812.2246",
    archivePrefix = "arXiv",
    primaryClass = "hep-th",
    doi = "10.1088/1126-6708/2009/04/118",
    journal = "JHEP",
    volume = "04",
    pages = "118",
    year = "2009"
}

@article{Dubovsky:2011uy,
    author = "Dubovsky, Sergei and Senatore, Leonardo and Villadoro, Giovanni",
    title = "{Universality of the Volume Bound in Slow-Roll Eternal Inflation}",
    eprint = "1111.1725",
    archivePrefix = "arXiv",
    primaryClass = "hep-th",
    reportNumber = "SLAC-PUB-14663",
    doi = "10.1007/JHEP05(2012)035",
    journal = "JHEP",
    volume = "05",
    pages = "035",
    year = "2012"
}

@article{Sekino:2008he,
    author = "Sekino, Yasuhiro and Susskind, Leonard",
    title = "{Fast Scramblers}",
    eprint = "0808.2096",
    archivePrefix = "arXiv",
    primaryClass = "hep-th",
    reportNumber = "SU-ITP-08-18, OIQP-08-08, SU-ITP-08/18, OIQP-08-08",
    doi = "10.1088/1126-6708/2008/10/065",
    journal = "JHEP",
    volume = "10",
    pages = "065",
    year = "2008"
}

@article{Bedroya:2020rmd,
    author = "Bedroya, Alek",
    title = "{de Sitter Complementarity, TCC, and the Swampland}",
    eprint = "2010.09760",
    archivePrefix = "arXiv",
    primaryClass = "hep-th",
    doi = "10.31526/lhep.2021.187",
    journal = "LHEP",
    volume = "2021",
    pages = "187",
    year = "2021"
}

@article{Aalsma:2020aib,
    author = "Aalsma, Lars and Shiu, Gary",
    title = "{Chaos and complementarity in de Sitter space}",
    eprint = "2002.01326",
    archivePrefix = "arXiv",
    primaryClass = "hep-th",
    reportNumber = "MAD-TH-20-01",
    doi = "10.1007/JHEP05(2020)152",
    journal = "JHEP",
    volume = "05",
    pages = "152",
    year = "2020"
}

@article{Cui:2026bcd,
    author = "Cui, Wentao and Kolchmeyer, David K.",
    title = "{A de Sitter Anti-Scrambling Algebra}",
    eprint = "2607.13665",
    archivePrefix = "arXiv",
    primaryClass = "hep-th",
    reportNumber = "MIT-CTP/6067",
    month = "7",
    year = "2026"
}

@article{Harlow:2026pwe,
    author = "Harlow, Daniel and Zhao, Ying",
    title = "{Anti-scrambling and euclidean folds from observer correlators in de Sitter space}",
    eprint = "2607.14215",
    archivePrefix = "arXiv",
    primaryClass = "hep-th",
    month = "7",
    year = "2026"
}

@article{Bedroya:2019tba,
    author = "Bedroya, Alek and Brandenberger, Robert and Loverde, Marilena and Vafa, Cumrun",
    title = "{Trans-Planckian Censorship and Inflationary Cosmology}",
    eprint = "1909.11106",
    archivePrefix = "arXiv",
    primaryClass = "hep-th",
    doi = "10.1103/PhysRevD.101.103502",
    journal = "Phys. Rev. D",
    volume = "101",
    number = "10",
    pages = "103502",
    year = "2020"
}

@article{Guth:1982pn,
    author = "Guth, Alan H. and Weinberg, Erick J.",
    title = "{Could the Universe Have Recovered from a Slow First Order Phase Transition?}",
    reportNumber = "MIT-CTP-950",
    doi = "10.1016/0550-3213(83)90307-3",
    journal = "Nucl. Phys. B",
    volume = "212",
    pages = "321--364",
    year = "1983"
}

@article{Bedroya:2020rac,
    author = "Bedroya, Alek and Montero, Miguel and Vafa, Cumrun and Valenzuela, Irene",
    title = "{de Sitter Bubbles and the Swampland}",
    eprint = "2008.07555",
    archivePrefix = "arXiv",
    primaryClass = "hep-th",
    doi = "10.1002/prop.202000084",
    journal = "Fortsch. Phys.",
    volume = "68",
    number = "11-12",
    pages = "2000084",
    year = "2020"
}

@article{Creminelli:2008es,
    author = "Creminelli, Paolo and Dubovsky, Sergei and Nicolis, Alberto and Senatore, Leonardo and Zaldarriaga, Matias",
    title = "{The Phase Transition to Slow-roll Eternal Inflation}",
    eprint = "0802.1067",
    archivePrefix = "arXiv",
    primaryClass = "hep-th",
    reportNumber = "IC-2008-002",
    doi = "10.1088/1126-6708/2008/09/036",
    journal = "JHEP",
    volume = "09",
    pages = "036",
    year = "2008"
}

@article{Rudelius:2019cfh,
    author = "Rudelius, Tom",
    title = "{Conditions for (No) Eternal Inflation}",
    eprint = "1905.05198",
    archivePrefix = "arXiv",
    primaryClass = "hep-th",
    doi = "10.1088/1475-7516/2019/08/009",
    journal = "JCAP",
    volume = "08",
    pages = "009",
    year = "2019"
}

@article{Gorbenko:2019rza,
    author = "Gorbenko, Victor and Senatore, Leonardo",
    title = "{$\lambda \phi^4$ in dS}",
    eprint = "1911.00022",
    archivePrefix = "arXiv",
    primaryClass = "hep-th",
    month = "10",
    year = "2019"
}

@article{Mirbabayi:2019qtx,
    author = "Mirbabayi, Mehrdad",
    title = "{Infrared dynamics of a light scalar field in de Sitter}",
    eprint = "1911.00564",
    archivePrefix = "arXiv",
    primaryClass = "hep-th",
    doi = "10.1088/1475-7516/2020/12/006",
    journal = "JCAP",
    volume = "12",
    pages = "006",
    year = "2020"
}

@article{Baumgart:2019clc,
    author = "Baumgart, Matthew and Sundrum, Raman",
    title = "{De Sitter Diagrammar and the Resummation of Time}",
    eprint = "1912.09502",
    archivePrefix = "arXiv",
    primaryClass = "hep-th",
    doi = "10.1007/JHEP07(2020)119",
    journal = "JHEP",
    volume = "07",
    pages = "119",
    year = "2020"
}

@article{Cohen:2020php,
    author = "Cohen, Timothy and Green, Daniel",
    title = "{Soft de Sitter Effective Theory}",
    eprint = "2007.03693",
    archivePrefix = "arXiv",
    primaryClass = "hep-th",
    doi = "10.1007/JHEP12(2020)041",
    journal = "JHEP",
    volume = "12",
    pages = "041",
    year = "2020"
}

@article{Green:2026nnw,
    author = "Green, Daniel and Gupta, Kshitij and Premkumar, Akhil",
    title = "{The Quantum Mechanics of Rare Events: From Quantum Walks to Stochastic Inflation}",
    eprint = "2608.06319",
    archivePrefix = "arXiv",
    primaryClass = "hep-th",
    month = "8",
    year = "2026"
}

@article{Liddle:2003as,
    author = "Liddle, Andrew R and Leach, Samuel M",
    title = "{How long before the end of inflation were observable perturbations produced?}",
    eprint = "astro-ph/0305263",
    archivePrefix = "arXiv",
    doi = "10.1103/PhysRevD.68.103503",
    journal = "Phys. Rev. D",
    volume = "68",
    pages = "103503",
    year = "2003"
}

@article{deSalas:2015glj,
    author = "de Salas, P. F. and Lattanzi, M. and Mangano, G. and Miele, G. and Pastor, S. and Pisanti, O.",
    title = "{Bounds on very low reheating scenarios after Planck}",
    eprint = "1511.00672",
    archivePrefix = "arXiv",
    primaryClass = "astro-ph.CO",
    reportNumber = "IFIC-15-70",
    doi = "10.1103/PhysRevD.92.123534",
    journal = "Phys. Rev. D",
    volume = "92",
    number = "12",
    pages = "123534",
    year = "2015"
}

@article{Lyth:1996kt,
    author = "Lyth, David H. and Stewart, Ewan D.",
    title = "{More varieties of hybrid inflation}",
    eprint = "hep-ph/9606412",
    archivePrefix = "arXiv",
    reportNumber = "LANCS-TH-96-13, RESCEU-20-96",
    doi = "10.1103/PhysRevD.54.7186",
    journal = "Phys. Rev. D",
    volume = "54",
    pages = "7186--7190",
    year = "1996"
}

@article{Planck:2018jri,
    author = "Akrami, Y. and others",
    collaboration = "Planck",
    title = "{Planck 2018 results. X. Constraints on inflation}",
    eprint = "1807.06211",
    archivePrefix = "arXiv",
    primaryClass = "astro-ph.CO",
    doi = "10.1051/0004-6361/201833887",
    journal = "Astron. Astrophys.",
    volume = "641",
    pages = "A10",
    year = "2020"
}

@article{AtacamaCosmologyTelescope:2025blo,
    author = "Louis, Thibaut and others",
    collaboration = "Atacama Cosmology Telescope",
    title = "{The Atacama Cosmology Telescope: DR6 power spectra, likelihoods and {\ensuremath{\Lambda}}CDM parameters}",
    eprint = "2503.14452",
    archivePrefix = "arXiv",
    primaryClass = "astro-ph.CO",
    reportNumber = "FERMILAB-PUB-25-0071-PPD",
    doi = "10.1088/1475-7516/2025/11/062",
    journal = "JCAP",
    volume = "11",
    pages = "062",
    year = "2025"
}

@article{Gibbons:1977mu,
    author = "Gibbons, G. W. and Hawking, S. W.",
    title = "{Cosmological Event Horizons, Thermodynamics, and Particle Creation}",
    doi = "10.1103/PhysRevD.15.2738",
    journal = "Phys. Rev. D",
    volume = "15",
    pages = "2738--2751",
    year = "1977"
}

@article{Mizuno:2019bxy,
    author = "Mizuno, Shuntaro and Mukohyama, Shinji and Pi, Shi and Zhang, Yun-Long",
    title = "{Universal Upper Bound on the Inflationary Energy Scale from the Trans-Planckian Censorship Conjecture}",
    eprint = "1910.02979",
    archivePrefix = "arXiv",
    primaryClass = "astro-ph.CO",
    reportNumber = "IPMU19-0138, YITP-19-91",
    doi = "10.1103/PhysRevD.102.021301",
    journal = "Phys. Rev. D",
    volume = "102",
    number = "2",
    pages = "021301",
    year = "2020"
}

@article{Kamali:2019gzr,
    author = "Kamali, Vahid and Brandenberger, Robert",
    title = "{Relaxing the TCC Bound on Inflationary Cosmology?}",
    eprint = "2001.00040",
    archivePrefix = "arXiv",
    primaryClass = "hep-th",
    doi = "10.1140/epjc/s10052-020-7908-8",
    journal = "Eur. Phys. J. C",
    volume = "80",
    number = "4",
    pages = "339",
    year = "2020"
}

@article{Brandenberger:2019eni,
    author = "Brandenberger, Robert and Wilson-Ewing, Edward",
    title = "{Strengthening the TCC Bound on Inflationary Cosmology}",
    eprint = "2001.00043",
    archivePrefix = "arXiv",
    primaryClass = "hep-th",
    doi = "10.1088/1475-7516/2020/03/047",
    journal = "JCAP",
    volume = "03",
    pages = "047",
    year = "2020"
}

@article{Vilenkin:1983xq,
    author = "Vilenkin, Alexander",
    title = "{The Birth of Inflationary Universes}",
    reportNumber = "TUTP-83-1",
    doi = "10.1103/PhysRevD.27.2848",
    journal = "Phys. Rev. D",
    volume = "27",
    pages = "2848",
    year = "1983"
}

@article{Linde:1986fd,
    author = "Linde, Andrei D.",
    title = "{Eternally Existing Selfreproducing Chaotic Inflationary Universe}",
    reportNumber = "Print-86-0418 (LEBEDEV INST), LEBEDEV-86-106",
    doi = "10.1016/0370-2693(86)90611-8",
    journal = "Phys. Lett. B",
    volume = "175",
    pages = "395--400",
    year = "1986"
}

@article{Obied:2018sgi,
    author = "Obied, Georges and Ooguri, Hirosi and Spodyneiko, Lev and Vafa, Cumrun",
    title = "{De Sitter Space and the Swampland}",
    eprint = "1806.08362",
    archivePrefix = "arXiv",
    primaryClass = "hep-th",
    reportNumber = "CALT-TH-2018-020, IPMU18-0100",
    month = "6",
    year = "2018"
}

@article{Ooguri:2018wrx,
    author = "Ooguri, Hirosi and Palti, Eran and Shiu, Gary and Vafa, Cumrun",
    title = "{Distance and de Sitter Conjectures on the Swampland}",
    eprint = "1810.05506",
    archivePrefix = "arXiv",
    primaryClass = "hep-th",
    doi = "10.1016/j.physletb.2018.11.018",
    journal = "Phys. Lett. B",
    volume = "788",
    pages = "180--184",
    year = "2019"
}

@article{Matsui:2018bsy,
    author = "Matsui, Hiroki and Takahashi, Fuminobu",
    title = "{Eternal Inflation and Swampland Conjectures}",
    eprint = "1807.11938",
    archivePrefix = "arXiv",
    primaryClass = "hep-th",
    reportNumber = "TU-1068, IPMU18-0132, MIT-CTP/5040",
    doi = "10.1103/PhysRevD.99.023533",
    journal = "Phys. Rev. D",
    volume = "99",
    number = "2",
    pages = "023533",
    year = "2019"
}

@article{Dimopoulos:2018upl,
    author = "Dimopoulos, Konstantinos",
    title = "{Steep Eternal Inflation and the Swampland}",
    eprint = "1810.03438",
    archivePrefix = "arXiv",
    primaryClass = "gr-qc",
    doi = "10.1103/PhysRevD.98.123516",
    journal = "Phys. Rev. D",
    volume = "98",
    number = "12",
    pages = "123516",
    year = "2018"
}

@article{Kinney:2018kew,
    author = "Kinney, William H.",
    title = "{Eternal Inflation and the Refined Swampland Conjecture}",
    eprint = "1811.11698",
    archivePrefix = "arXiv",
    primaryClass = "astro-ph.CO",
    doi = "10.1103/PhysRevLett.122.081302",
    journal = "Phys. Rev. Lett.",
    volume = "122",
    number = "8",
    pages = "081302",
    year = "2019"
}

@article{Brahma:2019iyy,
    author = "Brahma, Suddhasattwa and Shandera, Sarah",
    title = "{Stochastic eternal inflation is in the swampland}",
    eprint = "1904.10979",
    archivePrefix = "arXiv",
    primaryClass = "hep-th",
    doi = "10.1007/JHEP11(2019)016",
    journal = "JHEP",
    volume = "11",
    pages = "016",
    year = "2019"
}

@article{Wang:2019eym,
    author = "Wang, Ziwei and Brandenberger, Robert and Heisenberg, Lavinia",
    title = "{Eternal Inflation, Entropy Bounds and the Swampland}",
    eprint = "1907.08943",
    archivePrefix = "arXiv",
    primaryClass = "hep-th",
    doi = "10.1140/epjc/s10052-020-8412-x",
    journal = "Eur. Phys. J. C",
    volume = "80",
    number = "9",
    pages = "864",
    year = "2020"
}

@article{Blanco-Pillado:2019tdf,
    author = "Blanco-Pillado, Jose J. and Deng, Heling and Vilenkin, Alexander",
    title = "{Eternal Inflation in Swampy Landscapes}",
    eprint = "1909.00068",
    archivePrefix = "arXiv",
    primaryClass = "gr-qc",
    doi = "10.1088/1475-7516/2020/05/014",
    journal = "JCAP",
    volume = "05",
    pages = "014",
    year = "2020"
}

@article{Martin:2000xs,
    author = "Martin, Jerome and Brandenberger, Robert H.",
    title = "{The TransPlanckian problem of inflationary cosmology}",
    eprint = "hep-th/0005209",
    archivePrefix = "arXiv",
    doi = "10.1103/PhysRevD.63.123501",
    journal = "Phys. Rev. D",
    volume = "63",
    pages = "123501",
    year = "2001"
}

@article{Bedroya:2022tbh,
    author = "Bedroya, Alek",
    title = "{Holographic origin of TCC and the distance conjecture}",
    eprint = "2211.09128",
    archivePrefix = "arXiv",
    primaryClass = "hep-th",
    doi = "10.1007/JHEP06(2024)016",
    journal = "JHEP",
    volume = "06",
    pages = "016",
    year = "2024"
}

@article{Bedroya:2025fie,
    author = "Bedroya, Alek and Steinhardt, Paul J.",
    title = "{Holographic Constraints on the String Landscape}",
    eprint = "2511.15784",
    archivePrefix = "arXiv",
    primaryClass = "hep-th",
    month = "11",
    year = "2025"
}

@article{Bedroya:2024zta,
    author = "Bedroya, Alek and Lu, Qianshu and Steinhardt, Paul J.",
    title = "{TCC in the interior of moduli space and its implications for the string landscape and cosmology}",
    eprint = "2407.08793",
    archivePrefix = "arXiv",
    primaryClass = "hep-th",
    doi = "10.1007/JHEP08(2025)007",
    journal = "JHEP",
    volume = "08",
    pages = "007",
    year = "2025"
}

@article{Saito:2019tkc,
    author = "Saito, Ryo and Shirai, Satoshi and Yamazaki, Masahito",
    title = "{Is the trans-Planckian censorship a swampland conjecture?}",
    eprint = "1911.10445",
    archivePrefix = "arXiv",
    primaryClass = "hep-th",
    reportNumber = "IPMU-19-0170",
    doi = "10.1103/PhysRevD.101.046022",
    journal = "Phys. Rev. D",
    volume = "101",
    number = "4",
    pages = "046022",
    year = "2020"
}

@article{Burgess:2020nec,
    author = "Burgess, C. P. and de Alwis, S. P. and Quevedo, F.",
    title = "{Cosmological Trans-Planckian Conjectures are not Effective}",
    eprint = "2011.03069",
    archivePrefix = "arXiv",
    primaryClass = "hep-th",
    doi = "10.1088/1475-7516/2021/05/037",
    journal = "JCAP",
    volume = "05",
    pages = "037",
    year = "2021"
}

@article{Dvali:2020cgt,
    author = "Dvali, Gia and Kehagias, Alex and Riotto, Antonio",
    title = "{Inflation and Decoupling}",
    eprint = "2005.05146",
    archivePrefix = "arXiv",
    primaryClass = "hep-th",
    month = "5",
    year = "2020"
}

@article{Starobinsky:1986fx,
    author = "Starobinsky, Alexei A.",
    title = "{STOCHASTIC DE SITTER (INFLATIONARY) STAGE IN THE EARLY UNIVERSE}",
    doi = "10.1007/3-540-16452-9_6",
    journal = "Lect. Notes Phys.",
    volume = "246",
    pages = "107--126",
    year = "1986"
}

@article{Starobinsky:1994bd,
    author = "Starobinsky, Alexei A. and Yokoyama, Junichi",
    title = "{Equilibrium state of a selfinteracting scalar field in the De Sitter background}",
    eprint = "astro-ph/9407016",
    archivePrefix = "arXiv",
    reportNumber = "YITP-U-94-12",
    doi = "10.1103/PhysRevD.50.6357",
    journal = "Phys. Rev. D",
    volume = "50",
    pages = "6357--6368",
    year = "1994"
}

@article{Vennin:2015hra,
    author = "Vennin, Vincent and Starobinsky, Alexei A.",
    title = "{Correlation Functions in Stochastic Inflation}",
    eprint = "1506.04732",
    archivePrefix = "arXiv",
    primaryClass = "hep-th",
    doi = "10.1140/epjc/s10052-015-3643-y",
    journal = "Eur. Phys. J. C",
    volume = "75",
    pages = "413",
    year = "2015"
}

@article{Pattison:2017mbe,
    author = "Pattison, Chris and Vennin, Vincent and Assadullahi, Hooshyar and Wands, David",
    title = "{Quantum diffusion during inflation and primordial black holes}",
    eprint = "1707.00537",
    archivePrefix = "arXiv",
    primaryClass = "hep-th",
    doi = "10.1088/1475-7516/2017/10/046",
    journal = "JCAP",
    volume = "10",
    pages = "046",
    year = "2017"
}

@article{Ezquiaga:2019ftu,
    author = "Ezquiaga, Jose Mar{\'\i}a and Garc{\'\i}a-Bellido, Juan and Vennin, Vincent",
    title = "{The exponential tail of inflationary fluctuations: consequences for primordial black holes}",
    eprint = "1912.05399",
    archivePrefix = "arXiv",
    primaryClass = "astro-ph.CO",
    doi = "10.1088/1475-7516/2020/03/029",
    journal = "JCAP",
    volume = "03",
    pages = "029",
    year = "2020"
}

@article{Anchordoqui:2026edd,
    author = "Anchordoqui, Luis A. and Antoniadis, Ignatios and Bedroya, Alek",
    title = "{Distance-Higuchi Bounds on Inflationary Field Ranges and Lifetimes}",
    eprint = "2608.11086",
    archivePrefix = "arXiv",
    primaryClass = "hep-th",
    month = "8",
    year = "2026"
}

@article{Guleryuz:2026rdz,
    author = "Guleryuz, Omer",
    title = "{Stochastic Survival near Swampland Boundaries}",
    eprint = "2606.08244",
    archivePrefix = "arXiv",
    primaryClass = "hep-th",
    month = "6",
    year = "2026"
}

@article{Winitzki:2008ph,
    author = "Winitzki, Sergei",
    title = "{Reheating-volume measure for random-walk inflation}",
    eprint = "0805.3940",
    archivePrefix = "arXiv",
    primaryClass = "gr-qc",
    reportNumber = "YITP-08-41",
    doi = "10.1103/PhysRevD.78.063517",
    journal = "Phys. Rev. D",
    volume = "78",
    pages = "063517",
    year = "2008"
}

@article{Animali:2024jiz,
    author = "Animali, Chiara and Vennin, Vincent",
    title = "{Clustering of primordial black holes from quantum diffusion during inflation}",
    eprint = "2402.08642",
    archivePrefix = "arXiv",
    primaryClass = "astro-ph.CO",
    doi = "10.1088/1475-7516/2024/08/026",
    journal = "JCAP",
    volume = "08",
    pages = "026",
    year = "2024"
}

@article{Harlow:2011az,
    author = "Harlow, Daniel and Shenker, Stephen H. and Stanford, Douglas and Susskind, Leonard",
    title = "{Tree-like structure of eternal inflation: A solvable model}",
    eprint = "1110.0496",
    archivePrefix = "arXiv",
    primaryClass = "hep-th",
    reportNumber = "SU-ITP-11-47",
    doi = "10.1103/PhysRevD.85.063516",
    journal = "Phys. Rev. D",
    volume = "85",
    pages = "063516",
    year = "2012"
}

@article{KESTEN19789,
title = {Branching brownian motion with absorption},
journal = {Stochastic Processes and their Applications},
volume = {7},
number = {1},
pages = {9-47},
year = {1978},
issn = {0304-4149},
doi = {https://doi.org/10.1016/0304-4149(78)90035-2},
url = {https://www.sciencedirect.com/science/article/pii/0304414978900352},
author = {Harry Kesten}
}

@article{10.1214/ECP.v12-1259,
author = {John Harris and Simon Harris},
title = {{Survival probabilities for branching Brownian motion with absorption}},
volume = {12},
journal = {Electronic Communications in Probability},
number = {none},
publisher = {Institute of Mathematical Statistics and Bernoulli Society},
pages = {81 -- 92},
year = {2007},
doi = {10.1214/ECP.v12-1259},
URL = {https://doi.org/10.1214/ECP.v12-1259}
}

@article{berestycki2014critical,
  title={Critical branching Brownian motion with absorption: survival probability},
  eprint = "1212.3821",
  archivePrefix = "arXiv",
  primaryClass={math.PR},
  author={Berestycki, Julien and Berestycki, Nathana{\"e}l and Schweinsberg, Jason},
  doi="10.1007/s00440-013-0533-9",
  journal={Probability Theory and Related Fields},
  volume={160},
  number={3},
  pages={489--520},
  year={2014},
  publisher={Springer},
}

@article{AHL_2022__5__921_0,
     author = {Maillard, Pascal and Schweinsberg, Jason},
     title = {Yaglom-type limit theorems for branching {Brownian} motion with absorption},
     eprint="2010.16133", 
     archivePrefix="arXiv", primaryClass="math.PR",
     journal = {Annales Henri Lebesgue},
     pages = {921--985},
     year = {2022},
     publisher = {\'ENS Rennes},
     volume = {5},
     doi = {10.5802/ahl.140},
     language = {en},
     url = {https://ahl.centre-mersenne.org/articles/10.5802/ahl.140/}
}

@article{Fujita:2013cna,
    author = "Fujita, Tomohiro and Kawasaki, Masahiro and Tada, Yuichiro and Takesako, Tomohiro",
    title = "{A new algorithm for calculating the curvature perturbations in stochastic inflation}",
    eprint = "1308.4754",
    archivePrefix = "arXiv",
    primaryClass = "astro-ph.CO",
    reportNumber = "IPMU-13-0154, ICRR-REPORT-658-2013-7",
    doi = "10.1088/1475-7516/2013/12/036",
    journal = "JCAP",
    volume = "12",
    pages = "036",
    year = "2013"
}

@article{Winitzki:2001np,
    author = "Winitzki, Serge",
    title = "{The Eternal fractal in the universe}",
    eprint = "gr-qc/0111048",
    archivePrefix = "arXiv",
    doi = "10.1103/PhysRevD.65.083506",
    journal = "Phys. Rev. D",
    volume = "65",
    pages = "083506",
    year = "2002"
}

@article{Berestycki_2011,
   title={Survival of Near-Critical Branching Brownian Motion},
   eprint="1009.0406", archivePrefix="arXiv", primaryClass="math.PR",
   volume={143},
   ISSN={1572-9613},
   url={http://dx.doi.org/10.1007/s10955-011-0224-9},
   DOI={10.1007/s10955-011-0224-9},
   number={5},
   journal={Journal of Statistical Physics},
   publisher={Springer Science and Business Media LLC},
   author={Berestycki, Julien and Berestycki, Nathanaël and Schweinsberg, Jason},
   year={2011},
   month=May, pages={833–854} }

@article{Derrida_2007,
   title={The survival probability of a branching random walk in presence of an absorbing wall},
   eprint="cond-mat/0703353", archivePrefix="arXiv", 
   primaryClass="cond-mat.stat-mech",
   volume={78},
   ISSN={1286-4854},
   url={http://dx.doi.org/10.1209/0295-5075/78/60006},
   DOI={10.1209/0295-5075/78/60006},
   number={6},
   journal={Europhysics Letters (EPL)},
   publisher={IOP Publishing},
   author={Derrida, B and Simon, D},
   year={2007},
   pages={60006} }

@article{DeSimone:2008bq,
    author = "De Simone, Andrea and Guth, Alan H. and Salem, Michael P. and Vilenkin, Alexander",
    title = "{Predicting the cosmological constant with the scale-factor cutoff measure}",
    eprint = "0805.2173",
    archivePrefix = "arXiv",
    primaryClass = "hep-th",
    reportNumber = "MIT-CTP-3934",
    doi = "10.1103/PhysRevD.78.063520",
    journal = "Phys. Rev. D",
    volume = "78",
    pages = "063520",
    year = "2008"
}

@article{DeSimone:2008if,
    author = "De Simone, Andrea and Guth, Alan H. and Linde, Andrei D. and Noorbala, Mahdiyar and Salem, Michael P. and Vilenkin, Alexander",
    title = "{Boltzmann brains and the scale-factor cutoff measure of the multiverse}",
    eprint = "0808.3778",
    archivePrefix = "arXiv",
    primaryClass = "hep-th",
    reportNumber = "MIT-CTP-3975, SU-ITP-08-20",
    doi = "10.1103/PhysRevD.82.063520",
    journal = "Phys. Rev. D",
    volume = "82",
    pages = "063520",
    year = "2010"
}

@article{Bousso:2008hz,
    author = "Bousso, Raphael and Freivogel, Ben and Yang, I-Sheng",
    title = "{Properties of the scale factor measure}",
    eprint = "0808.3770",
    archivePrefix = "arXiv",
    primaryClass = "hep-th",
    doi = "10.1103/PhysRevD.79.063513",
    journal = "Phys. Rev. D",
    volume = "79",
    pages = "063513",
    year = "2009"
}

@article{Winitzki:2008yb,
    author = "Winitzki, Sergei",
    title = "{A Volume-weighted measure for eternal inflation}",
    eprint = "0803.1300",
    archivePrefix = "arXiv",
    primaryClass = "gr-qc",
    doi = "10.1103/PhysRevD.78.043501",
    journal = "Phys. Rev. D",
    volume = "78",
    pages = "043501",
    year = "2008"
}

@article{Cotler:2026lna,
    author = "Cotler, Jordan and Ivo, Victor and Maldacena, Juan",
    title = "{Renormalization in Liouville gravity and stochastic inflation}",
    eprint = "2608.26244",
    archivePrefix = "arXiv",
    primaryClass = "hep-th",
    month = "8",
    year = "2026"
}

@article{Cruces:2022imf,
    author = "Cruces, Diego",
    title = "{Review on Stochastic Approach to Inflation}",
    eprint = "2203.13852",
    archivePrefix = "arXiv",
    primaryClass = "gr-qc",
    doi = "10.3390/universe8060334",
    journal = "Universe",
    volume = "8",
    number = "6",
    pages = "334",
    year = "2022"
}

@article{hardy2006newformulationspineapproach,
      title={A new formulation of the spine approach to branching diffusions}, 
      author={Robert Hardy and Simon C. Harris},
      year={2006},
      eprint={math/0611054},
      archivePrefix={arXiv},
      primaryClass={math.PR},
      url={https://arxiv.org/abs/math/0611054}, 
}

@article{10.1214/15-AIHP714,
author = {Simon C. Harris and Matthew I. Roberts},
title = {{The many-to-few lemma and multiple spines}},
eprint="1106.4761", archivePrefix="arXiv", primaryClass="math.PR",
volume = {53},
journal = {Annales de l'Institut Henri Poincaré, Probabilités et Statistiques},
number = {1},
publisher = {Institut Henri Poincaré},
pages = {226 -- 242},
year = {2017},
doi = {10.1214/15-AIHP714},
URL = {https://doi.org/10.1214/15-AIHP714}
}

@article{Bousso:2006ev,
    author = "Bousso, Raphael",
    title = "{Holographic probabilities in eternal inflation}",
    eprint = "hep-th/0605263",
    archivePrefix = "arXiv",
    reportNumber = "UCB-PTH-06-10, LBNL-60251",
    doi = "10.1103/PhysRevLett.97.191302",
    journal = "Phys. Rev. Lett.",
    volume = "97",
    pages = "191302",
    year = "2006"
}

@article{Garriga:2005av,
    author = "Garriga, Jaume and Schwartz-Perlov, Delia and Vilenkin, Alexander and Winitzki, Sergei",
    title = "{Probabilities in the inflationary multiverse}",
    eprint = "hep-th/0509184",
    archivePrefix = "arXiv",
    doi = "10.1088/1475-7516/2006/01/017",
    journal = "JCAP",
    volume = "01",
    pages = "017",
    year = "2006"
}

@article{Bousso:2009mw,
    author = "Bousso, Raphael and Yang, I-Sheng",
    title = "{Global-Local Duality in Eternal Inflation}",
    eprint = "0904.2386",
    archivePrefix = "arXiv",
    primaryClass = "hep-th",
    doi = "10.1103/PhysRevD.80.124024",
    journal = "Phys. Rev. D",
    volume = "80",
    pages = "124024",
    year = "2009"
}

@article{Tomberg:2025fku,
    author = "Tomberg, Eemeli and Dimopoulos, Konstantinos",
    title = "{Eternal inflation near inflection points: a challenge to primordial black hole models}",
    eprint = "2507.15522",
    archivePrefix = "arXiv",
    primaryClass = "astro-ph.CO",
    doi = "10.1088/1475-7516/2026/01/033",
    journal = "JCAP",
    volume = "01",
    pages = "033",
    year = "2026"
}

@article{Garriga:2015fdk,
    author = "Garriga, Jaume and Vilenkin, Alexander and Zhang, Jun",
    title = "{Black holes and the multiverse}",
    eprint = "1512.01819",
    archivePrefix = "arXiv",
    primaryClass = "hep-th",
    doi = "10.1088/1475-7516/2016/02/064",
    journal = "JCAP",
    volume = "02",
    pages = "064",
    year = "2016"
}

@article{Mckean1975ApplicationOB,
 title="{Application of Brownian motion to the equation of Kolmogorov-Petrovskii
 -Piskunov}",
  author={H. P. Mckean},
  journal={Communications on Pure and Applied Mathematics},
  year={1975},
  volume={28},
  pages={323-331},
  url={https://api.semanticscholar.org/CorpusID:119710406}
}

@article{VANSAARLOOS_2003,
   title={Front propagation into unstable states},
   eprint="cond-mat/0308540", archivePrefix="arXiv",
   volume={386},
   ISSN={0370-1573},
   url={http://dx.doi.org/10.1016/j.physrep.2003.08.001},
   DOI={10.1016/j.physrep.2003.08.001},
   number={2-6},
   journal={Physics Reports},
   publisher={Elsevier BV},
   author={VANSAARLOOS, W},
   year={2003},
   month=Nov, pages={29–222} }

@incollection{Vennin:2024yzl,
    author = "Vennin, Vincent and Wands, David",
    title = "{Quantum Diffusion and~Large Primordial Perturbations from~Inflation}",
    eprint = "2402.12672",
    archivePrefix = "arXiv",
    primaryClass = "astro-ph.CO",
    editor="Byrnes, Christian and others", booktitle="{Primordial Black Holes}", publisher="Springer", 
    doi="10.1007/978-981-97-8887-3_8", pages="201",
    year = "2025"
}

@article{Linde:1993xx,
    author = "Linde, Andrei D. and Linde, Dmitri A. and Mezhlumian, Arthur",
    title = "{From the Big Bang theory to the theory of a stationary universe}",
    eprint = "gr-qc/9306035",
    archivePrefix = "arXiv",
    reportNumber = "SU-ITP-93-13",
    doi = "10.1103/PhysRevD.49.1783",
    journal = "Phys. Rev. D",
    volume = "49",
    pages = "1783--1826",
    year = "1994"
}

@article{Vanchurin:1999iv,
    author = "Vanchurin, Vitaly and Vilenkin, Alexander and Winitzki, Serge",
    title = "{Predictability crisis in inflationary cosmology and its resolution}",
    eprint = "gr-qc/9905097",
    archivePrefix = "arXiv",
    doi = "10.1103/PhysRevD.61.083507",
    journal = "Phys. Rev. D",
    volume = "61",
    pages = "083507",
    year = "2000"
}

@article{Mizuguchi:2024kbl,
    author = "Mizuguchi, Yurino and Murata, Tomoaki and Tada, Yuichiro",
    title = "{STOLAS: STOchastic LAttice Simulation of cosmic inflation}",
    eprint = "2405.10692",
    archivePrefix = "arXiv",
    primaryClass = "astro-ph.CO",
    reportNumber = "RUP-24-10",
    doi = "10.1088/1475-7516/2024/12/050",
    journal = "JCAP",
    volume = "12",
    pages = "050",
    year = "2024"
}

@article{https://doi.org/10.1111/j.1469-1809.1937.tb02153.x,
author = {FISHER, R. A.},
title = {THE WAVE OF ADVANCE OF ADVANTAGEOUS GENES},
journal = {Annals of Eugenics},
volume = {7},
number = {4},
pages = {355-369},
doi="10.1111/j.1469-1809.1937.tb02153.x",
year = {1937}
}

@article{KPP1937,
  author = {Kolmogorov, A. and Petrovskii, I. and Piscunov, N.},
  journal = {Byul. Moskovskogo Gos. Univ.},
  number = 6,
  pages = {1--25},
  title = {A study of the equation of diffusion with increase in the quantity of matter, and its application to a biological problem},
  url = {http://books.google.com/books?id=ikN59GkYJKIC&lpg=PP1&dq=A.N.%20Kolmogorov%3A%20Selected%20Works&client=firefox-a&pg=PA242#v=onepage&q=&f=false},
  volume = 1,
  year = 1937
}

\end{document}